\documentclass[oldversion]{aa}

\usepackage{txfonts}
\usepackage{graphicx}
\usepackage{natbib}
\usepackage{fixltx2e}
\bibpunct{(}{)}{;}{a}{}{,} 

\newcommand{\HI}{{\sc H\,i}}

\begin{document}

\title{The Westerbork Hydrogen Accretion in Local Galaxies\\(HALOGAS) Survey}
\subtitle{I. Survey Description and Pilot Observations}

\author{George Heald \inst{1} \and 
        Gyula J{\'o}zsa \inst{1} \and 
        Paolo Serra \inst{1} \and 
        Laura Zschaechner \inst{2} \and 
        Richard Rand \inst{2} \and 
        Filippo Fraternali \inst{3} \and 
        Tom Oosterloo \inst{1,4} \and 
        Rene Walterbos \inst{5} \and 
        Eva J\"utte \inst{6} \and 
        Gianfranco Gentile \inst{7} 
        }
\offprints{G. Heald, \email{heald@astron.nl}}
\institute{Netherlands Institute for Radio Astronomy (ASTRON), Postbus 2, 7990 AA Dwingeloo, the Netherlands \and 
           University of New Mexico, 800 Yale Blvd, Albuquerque, NM, USA \and 
           Astronomy Department, Bologna University, Via Ranzani 1, 40127 Bologna, Italy \and 
           Kapteyn Astronomical Institute, Postbus 800, 9700 AV Groningen, the Netherlands \and 
           Department of Astronomy, New Mexico State University, P.O. Box 30001, MSC 4500, Las Cruces, NM 88003, USA \and 
           Astronomisches Institut der Ruhr-Universit\"at Bochum, Universit\"atsstr. 150, 44780 Bochum, Germany \and 
           Sterrenkundig Observatorium, Ghent University, Krijgslaan 281, S9, B-9000 Ghent, Belgium 
           }
\date{Received 15 October 2010; accepted 27 November 2010}

\abstract{
We introduce a new, very deep neutral hydrogen (\HI) survey being performed with the Westerbork Synthesis Radio Telescope (WSRT). The Westerbork Hydrogen Accretion in LOcal GAlaxieS (HALOGAS) Survey is producing an archive of some of the most sensitive \HI\ observations available, on the angular scales which are most useful for studying faint, diffuse gas in and around nearby galaxies. The survey data are being used to perform careful modeling of the galaxies, characterizing their gas content, morphology, and kinematics, with the primary goal of revealing the global characteristics of cold gas accretion onto spiral galaxies in the local Universe. In this paper, we describe the survey sample selection, the data acquisition, reduction, and analysis, and present the data products obtained during our pilot program, which consists of UGC~2082, NGC~672, NGC~925, and NGC~4565. The observations reveal a first glimpse of the picture that the full HALOGAS project aims to illuminate: the properties of accreting \HI\ in different types of spirals, and across a range of galactic environments. None of the pilot survey galaxies hosts an \HI\ halo of the scale of NGC~891, but all show varying indications of halo gas features. We compare the properties of detected features in the pilot survey galaxies with their global characteristics, and discuss similarities and differences with NGC~891 and NGC~2403.}


\titlerunning{The WSRT HALOGAS Survey}
\authorrunning{G.~Heald et al.}

\maketitle

\section{Survey Motivation}

It is becoming increasingly clear that gas accretion is a crucial piece of the puzzle of galaxy evolution. Accretion has been invoked as a possible solution to several problems in galactic astrophysics. A typical accretion rate of $\approx\,1\,M_{\odot}\,\mathrm{yr}^{-1}$ seems to be necessary in order for spiral galaxies to replenish gas consumed in the star formation process -- it has been noted that otherwise, galaxies would use up their gas on timescales shorter than their ages \citep[e.g.,][]{larson_etal_1980}. The constancy of the SFR in the solar neighborhood \citep{twarog_1980,binney_etal_2000} argues that this inflow of fresh gas should locally be fairly constant. Similar accretion rates of metal-free gas may be needed to match the chemical evolution of the disk \citep[e.g.,][]{lacey_fall_1985,schoenrich_binney_2009}. Accretion of fresh gas has also been invoked to explain the ``G-dwarf problem,'' i.e. the small number of local stars with low metallicity \citep{pagel_1997}, for instance by \citet{chiappini_etal_1997}. Large scale disturbances in galaxies, such as lopsidedness (both morphological and kinematical) and warps, may be the result of accretion \citep{bournaud_etal_2005,sancisi_etal_2008}. In their recent review of the observational evidence for cold accretion, \citet{sancisi_etal_2008} infer a global average observed neutral gas accretion rate onto galaxies of $\gtrsim0.2\,M_{\odot}\,\mathrm{yr}^{-1}$. Such accretion may take place in two ways: through the merging of gas-rich satellites, and ``pure'' gas accretion from the intergalactic medium (IGM).

On theoretical grounds, a mode of accretion referred to as ``cold'' should continue in galaxies in low density environments through $z=0$ \citep[e.g.,][]{keres_etal_2005}. Note that this usage of the word ``cold'' differs from the meaning typically assumed by observers: in the theoretical picture of accretion, cold gas is that which is never shock heated to the virial temperature of the halo as it falls into the galactic potential, reaching maximum temperatures below $10^5\,\mathrm{K}$ during the accretion process; observationally, cold gas means gas at temperatures up to nearly $10^4\,\mathrm{K}$. Despite this difference in nomenclature, it seems clear that a significant portion of the accreting gas in local galaxies is expected to be at relatively low temperature, and that this gas phase will be traced, at least in part, by \HI.

Indeed, deep observations of a few nearby spirals -- most notably the edge-on NGC~891 \citep{oosterloo_etal_2007}, the moderately-inclined NGC~2403 \citep{fraternali_etal_2002}, and the face-on NGC~6946 \citep{boomsma_etal_2008} -- have shown that large quantities of cold gas exist in their halos, at large vertical distances from the thin disk. In the case of NGC~891, observations with the Westerbork Synthesis Radio Telescope (WSRT) revealed a huge neutral gas halo \cite[e.g.][see their Figure 1]{oosterloo_etal_2007} which extends up to a vertical distance of 22 kpc, with massive filamentary structures far from the disk. In total, the \HI\ halo of NGC~891 contains almost 30\% of the galaxy's total \HI\ mass.

How can so much gas be located at such large vertical distances from the disk? Star formation-driven disk-halo flows, as described by galactic fountain \citep[e.g.,][]{bregman_1980} or chimney \citep{norman_ikeuchi_1989} models, are likely part of the explanation. The morphology of the extraplanar gas seems to be related to the properties of the underlying star forming disk \citep[at least in its ionized phase; e.g.,][]{rand_1996}, and whether ionized gas is found in the halo or not seems to depend on the energy density provided by star formation \citep{rossa_dettmar_2003a}.

On the other hand, it has been shown that the rotational velocity of the halo gas (both in its neutral and ionized phases) decreases linearly with increasing distance from the midplane \citep{fraternali_etal_2005,heald_etal_2006,kamphuis_etal_2007}. This kinematic ``lag'' -- which allows us to distinguish halo gas from disk gas in moderately inclined galaxies such as NGC~2403 -- is much steeper than can be accounted for with simple galactic fountain models \citep[the discrepancy is observed to be a factor of two or more;][]{heald_etal_2007}. A good deal of theoretical effort has been put into this problem, and a range of more sophisticated models to explain the kinematics of gaseous halos have been put forward (using $N$-body+smoothed particle hydrodynamic (SPH) simulations, \citet{kaufmann_etal_2006}; hydrostatic models, \citet{benjamin_2000,barnabe_etal_2006,marinacci_etal_2010a}; SPH including spiral arms and bars, \citet{struck_smith_2009}; and cloud-corona interaction, \citet{marinacci_etal_2010b}). Such efforts have shown how the steep lag in NGC~891 might occur. Additionally, a constant accretion of low-angular momentum gas has been invoked, in combination with the simple galactic fountain picture, to slow the halo gas (the required accretion rate would be $\approx3\,M_{\odot}\,\mathrm{yr}^{-1}$ in NGC~891 based on modeling work by \citet{fraternali_binney_2008} -- in this picture, 90\% of the halo gas originates from galactic fountain, and 10\% from accretion). Distinguishing between these possible models requires a larger sample of gaseous halos and measurements of their kinematics.

It has been pointed out that if the largest \HI\ filament in the halo of NGC~891 was created by disk star formation, an energy equivalent to $\sim\,10^5$ supernovae would be required. This suggests an external origin instead \citep{oosterloo_etal_2007}. A final piece of evidence suggestive of accretion in NGC~891 comes in the form of small, $10^6\,M_{\odot}$ clouds which are {\em counter-rotating} in the halo. A filament with a mass of $10^7\,M_{\odot}$ is found in NGC~2403 \citep{fraternali_etal_2002}, and a counter-rotating $\sim10^7\,M_{\odot}$ cloud is also detected in the extraplanar region of NGC~5746 \citep{rand_benjamin_2008}.

Tracers of ongoing accretion can also be found closer to home: many of the MW's high velocity clouds (HVCs) have low metallicity \citep{tripp_etal_2003}; together with the increasingly tight constraints on their distances \citep{wakker_etal_2007,wakker_etal_2008}, the evidence is consistent with the idea that at least part of the HVC population is infalling primordial gas. HVC analogs have also been detected in the outskirts of M31 \citep{thilker_etal_2004}. In both the MW and M31, stellar streams are observed via deep optical observations \citep{belokurov_etal_2007,ferguson_2007}. The \HI\ filament in the halo of NGC~891 may be the cold gas counterpart of this type of stream. It has been recently recognized that NGC~891 also has a giant stellar stream looping around the disk, pointing to recent interaction \citep{mouhcine_etal_2010}. Observing such features in \HI\ emission has the advantage that we directly obtain kinematical information.

Halo gas is also present in at least one low surface brightness (LSB) galaxy (NGC~4395) with very little ongoing star formation, and a $10^7\,M_{\odot}$ gas cloud in that system is likely in the process of being accreted \citep{heald_oosterloo_2008}. Moreover, evidence suggests the presence of a neutral gas halo in the superthin LSB UGC 7321 \citep{matthews_wood_2003}. On the other hand, there is no evidence for large amounts of \HI\ in the halo of the very massive NGC 5746 \citep{rand_benjamin_2008}.

Why do some galaxies have massive gaseous halos, but not others? Is the \HI\ halo of NGC~891 typical or exceptional? At present, the extremely limited number of sufficiently deep observations restricts our ability to draw general conclusions about the prevalence of accretion in the local Universe. A comprehensive investigation is needed, using a representative sample of spiral galaxies. Most of the existing deep observations were prompted by earlier shallow observations which already showed evidence for an exceptional halo population or other peculiarities. Moreover, the targeted galaxies were chosen individually, rather than as a part of a larger program to sample a wide range of galactic disk properties such as star formation rates, warps, and lopsidedness.

As a first attempt to provide such information, we have started the Westerbork Hydrogen Accretion in LOcal GAlaxieS (HALOGAS) Survey. With this project we have the goal to make a significant step forward in understanding the process of cold gas accretion in spiral galaxies, by investigating the {\it general phenomenon} for the first time.

The HALOGAS Survey observations are designed to detect and characterize neutral extraplanar gas by measuring its total mass, rotation speed (in edge-ons), radial inflow or outflow \citep[in inclined galaxies such as NGC~2403; see the work by][]{fraternali_etal_2002}, and distribution (smooth, filamentary, or clumpy?). Deriving the extraplanar gas kinematics is of crucial importance -- these parameters are key predictions of halo gas and accretion models. Comparisons between the extraplanar gas characteristics and the properties of the spiral disk should clarify the roles of star formation and accretion.

The HALOGAS data will form a publicly available repository of the deepest \HI\ observations of nearby spiral galaxies. Additional studies which will be enabled by this dataset are numerous, and include:
\begin{enumerate}
\item Search for background continuum source variability at low ($\sim1$ mJy) flux levels, over the full HALOGAS survey area (totalling $\sim6.2$ square degrees within the WSRT primary beam FWHM);
\item Detailed analyses of the gravitational potential by measuring rotation curves extending far into the faint outer disks, and additionally by using the combination of the gas dispersion and vertical scaleheight;
\item Studies of warps and lopsidedness in spiral galaxies; and
\item The details of the \HI\ cutoff at the edge of the gaseous disk, at sensitivity levels comparable to the requirement specified by e.g. \citet{corbelli_bandiera_2002} to examine the \ion{H}{i} -- \ion{H}{ii} transition.
\end{enumerate}

The survey observations are being completed at the WSRT. Our first semester of observations was designed as a small pilot program, and that pilot survey is the focus of this paper.

The paper is organized as follows. We describe the selection of the survey sample in \S\,\ref{section:sample}. The observational setup and data reduction path are outlined in \S\,\ref{section:observations}. Results from the HALOGAS Pilot Survey are presented in \S\,\ref{section:results}. We discuss the first results from the HALOGAS Survey in \S\,\ref{section:discussion}. Finally, multi-wavelength supplements to the HALOGAS Survey, and our data release plans, are discussed in \S\,\ref{section:multiwavelength} and \ref{section:datarelease}.

\section{The HALOGAS Sample}\label{section:sample}

To reach the sensitivity required to detect and characterize faint, diffuse \HI, with characteristic column densities of a few times $10^{19}\,\mathrm{cm^{-2}}$, each of the HALOGAS targets is being observed at the WSRT for a total of $10\times12$hr. We would like to have an unbiased look at the accretion process in the local Universe. From this perspective, non-detections of extraplanar gas are as interesting as the detections -- by restricting attention to galaxies already suspected to show clear signs of accretion, we would learn less about the ubiquity of the accretion process in the spiral population, or the relationship between galaxy parameters and the accretion process. We therefore deliberately avoided selecting or rejecting sample galaxies based on existing \HI\ observations, or other subjective judgements about how interesting a target might appear when considered as an individual target. The final sample of galaxies was, ultimately, chosen on the basis of a small number of basic and neutral selection criteria, each of which are designed to maximize our ability to detect and study the properties of any diffuse gas which might be present in the target systems.

To select the galaxies for our sample, we searched in the \citet{tully_1988} catalog, requiring the following characteristics:
\begin{enumerate}
\item[(i)] The sample is made up of barred and unbarred spirals with Hubble types between Sa and Sd.
\item[(ii)] Because the WSRT is an east-west interferometer, we require that the source declination $\delta>+25^{\circ}$ to ensure that the synthesized beamsize is $\lesssim30''$.
\item[(iii)] We only select nearby galaxies with distance $D<11\,\mathrm{Mpc}$, in order to observe at adequate physical resolution to resolve the morphology of the galaxies (for instance, the characteristic WSRT beamsize of $15''=0.7\,\mathrm{kpc}$ at a target distance of 10 Mpc).
\item[(iv)] In order to ensure that each galaxy is sampled by a reasonably large number of angular resolution elements, we require a large optical isophotal diameter at 25th magnitude $D_{25}>3'$. Typically, \HI\ disks are larger than $D_{25}$.
\item[(v)] Finally, we require that the systemic velocity is $>100\mathrm{km\,s^{-1}}$, in order to avoid confusion with MW \HI.
\end{enumerate}

Within these criteria, we selected all intermediately inclined ($50^{\circ}\leq\,i\,\leq75^{\circ}$) and edge-on ($i\,\geq85^{\circ}$) galaxies. Galaxies in the former category reveal connections between halo gas and star formation, and allow investigation of rotation and radial motions in the halo gas \citep[e.g.][]{fraternali_etal_2002}; edge-on galaxies are useful for learning about the extent of halos, and the vertical change in rotation speed of the halo gas \citep[e.g.][]{oosterloo_etal_2007}. Visual inspection was utilized only to discard galaxies with incorrectly catalogued inclination angles. We also excluded from our new observations two galaxies which meet our criteria, but which are already deeply observed in \HI\ (namely, NGC~891 and NGC~2403). We will incorporate the properties of these galaxies into our final statistical sample since they match the sample criteria described above. The final sample of 24 galaxies is listed in Table \ref{table:sample}. The columns are (1) the UGC ID; (2) NGC or Messier IDs if available; (3) the Hubble type; (4) the distance $d_\mathrm{T88}$ in Mpc from \citet{tully_1988}; (5) the adopted distance $d_\mathrm{best}$ in Mpc; (6) the method for determining $d_\mathrm{best}$; and our best estimate using the tabulated method; (7) the systemic velocity $v_{\mathrm{sys}}$ in km\,s$^{-1}$; (8) the inclination $i$; (9) the optical diameter $D_{25}$; (10) the blue magnitude $M_B$; (11) the rotational velocity $v_{\mathrm{rot}}$; (12) the star formation rate (SFR); and (13) references used in deriving the SFR values in column (12), see below for details. The UGC IDs of the four targets which have been observed in the pilot survey described in this paper are indicated with a bold font.

\begin{table*}
\caption{The Full HALOGAS Sample.}
\label{table:sample}
\centering
\begin{tabular}{cllcccccccccl}
\hline\hline
~ & ~ & ~ & $d_\mathrm{T88}$ & $d_\mathrm{best}$ & $d_\mathrm{best}$ & $v_{\mathrm{sys}}$ & $i$ & $D_{25}$ & $M_{B}$ & $v_{\mathrm{rot}}$ & SFR & SFR \\
UGC & Other IDs & Type & (Mpc) & (Mpc) & method & $\mathrm{(km\,s^{-1})}$ & $(^{\circ})$ & (arcmin) & (mag) & $\mathrm{(km\,s^{-1})}$ & ($M_{\odot}\,\mathrm{yr}^{-1}$) & refs \\
\hline
{\bf 1256}        & NGC~0672        & SBcd  & 7.5  &  7.6 & 2 & 425  & 70 & 6.4  & -18.65 & 130.7 & 0.22               & H1,I2\\
1831              & NGC~0891        & SAb   & 9.6  &  9.2 & 1 & 529  & 84 & 12.2 & -19.96 & 212.2 & 2.4                & H1,I4\\
{\bf 1913}        & NGC~0925        & SABd  & 9.4  &  9.1 & 1 & 554  & 54 & 11.3 & -19.66 & 102.4 & 0.83               & H1,I4\\
1983              & NGC~0949        & SAd   & 10.3 & 11.3 & 2 & 610  & 52 & 3.5  & -17.85 & 90.9  & 0.26               & H1,I1\\
{\bf 2082}        & $-$             & SAc   & 10.7 & 14.4 & 3 & 710  & 89 & 5.8  & -18.55 & 86.6  & 0.023              & $-$,I1\\
2137              & NGC~1003        & SAcd  & 10.7 & 11.6 & 3 & 626  & 67 & 6.3  & -18.61 & 95.5  & 0.34               & H1,I2\\
3918              & NGC~2403        & SAcd  & 4.2  &  3.2 & 1 & 132  & 62 & 23.8 & -19.68 & 121.9 & 1.0                & H1,I4\\
4278              & IC~2233         & SAd   & 10.6 & 13.6 & 3 & 565  & 90 & 4.3  & -17.45 & 79.2  & 0.11               & H1,I5\\
4284              & NGC~2541        & SAcd  & 10.6 & 12.0 & 1 & 553  & 67 & 7.2  & -18.37 & 92.1  & $<0.27^\mathrm{a}$ & H2,I2\\
5572              & NGC~3198        & SBc   & 10.8 & 14.5 & 1 & 660  & 71 & 8.8  & -19.62 & 148.2 & 0.61               & H2,I4\\
7045              & NGC~4062        & SAc   & 9.7  & 16.9 & 3 & 769  & 68 & 4.5  & -18.27 & 140.5 & 0.22               & H2,I2\\
7322              & NGC~4244        & SAcd  & 3.1  &  4.4 & 1 & 247  & 90 & 15.8 & -17.60 & 89.0  & 0.058              & H1,I5\\
7353              & NGC~4258 (M106) & SABbc & 6.8  &  7.6 & 1 & 449  & 71 & 17.1 & -20.59 & 208.0 & 1.4                & H1,I3\\
7377              & NGC~4274        & SBab  & 9.7  & 19.4 & 3 & 922  & 72 & 6.5  & -19.22 & 239.9 & 0.31               & H3,I2\\
7539              & NGC~4414        & SAc   & 9.7  & 17.8 & 1 & 720  & 50 & 4.5  & -19.12 & 224.7 & 1.3                & H2,I4\\
7591$^\mathrm{b}$ & NGC~4448        & SBab  & 9.7  &  9.7 & 4 & 693  & 71 & 3.8  & -18.43 & 221.6 & 0.056              & $-$,I2\\
7766              & NGC~4559        & SABcd & 9.7  &  7.9 & 3 & 816  & 69 & 11.3 & -20.07 & 113.4 & 1.1                & H4,I4\\
{\bf 7772}        & NGC~4565        & SAb   & 9.7  & 10.8 & 2 & 1228 & 90 & 16.2 & -20.34 & 244.9 & 0.54               & H5,I4\\
7774$^\mathrm{b}$ & $-$             & SAd   & 6.8  & 24.4 & 3 & 526  & 90 & 3.5  & -15.57 & 79.4  & 0.0074             & H1,I5\\
7865              & NGC~4631        & SBd   & 6.9  &  7.6 & 1 & 613  & 85 & 14.7 & -20.12 & 138.9 & 2.1                & H1,I4\\
8286              & NGC~5023        & SAc   & 6.0  &  6.6 & 1 & 400  & 90 & 6.8  & -17.29 & 80.3  & 0.032              & H1,I5\\
8334              & NGC~5055 (M63)  & SAbc  & 7.2  &  8.5 & 3 & 497  & 55 & 13.0 & -20.14 & 215.5 & 1.5                & H1,I4\\
8550              & NGC~5229        & SBc   & 6.4  &  5.1 & 2 & 365  & 90 & 3.5  & -15.82 & 57.3  & 0.014              & H1,I5\\
9179              & NGC~5585        & SABd  & 7.0  &  8.7 & 2 & 303  & 51 & 5.5  & -17.96 & 79.1  & 0.26               & H1,I1\\
\hline
\end{tabular}
\begin{list}{}{}
\item[$^{\mathrm{a}}$] The SFR value for NGC~2541 is strictly speaking an upper limit, because the IRAS 25$\mu$m flux is catalogued as a non-detection by \citet{moshir_etal_1990}.
\item[$^{\mathrm{b}}$] No \HI\ observing time granted yet.
\end{list}
\tablebib{(H1) \citet{kennicutt_etal_2008}; (H2) \citet{moustakas_kennicutt_2006}; (H3) \citet{hameed_devereux_2005}; (H4) \citet{kennicutt_etal_2009}; (H5) \citet{robitaille_etal_2007}; (I1) \citet{lisenfeld_etal_2007}; (I2) \citet{moshir_etal_1990}; (I3) \citet{rice_etal_1988}; (I4) \citet{sanders_etal_2003}; (I5) \citet{dale_etal_2009}.}
\end{table*}

The galaxy parameters listed in Table \ref{table:sample} are taken from \citet{tully_1988}, except for $v_{\mathrm{rot}}$, which is from HyperLEDA \citep{paturel_etal_2003}, the adopted distance $d_\mathrm{best}$, and the SFR, both of which are discussed below. Given the broad range of Hubble type, SFR, mass, and environmental parameters (see Table \ref{table:environment} and the discussion below) exhibited by nearby spiral galaxies, this sample size stikes an acceptable balance between the number of targets required to reasonably sample the parameter space, while still resulting in a total amount of observing time which does not become prohibitive.

One of the sample selection criteria is to select only galaxies closer than 11~Mpc. This selection is based on galaxy distances taken from \citet{tully_1988}, which we indicate as $d_\mathrm{T88}$. More recent distance estimates are available for all selected galaxies. Adopting the best one is important when measuring, e.g., mass and linear size of the detected \HI\ structures. Therefore, we searched the NASA/IPAC Extragalactic Database (NED)\footnote{The NASA/IPAC Extragalactic Database (NED) is operated by the Jet Propulsion Laboratory, California Institute of Technology, under contract with the National Aeronautics and Space Administration.} distance database to obtain the most accurate distance of each galaxy, which we indicate as $d_\mathrm{best}$. Table \ref{table:sample} lists both $d_\mathrm{T88}$ and $d_\mathrm{best}$ for all galaxies in the sample. We define $d_\mathrm{best}$ using one of the following methods, sorted according to decreasing priority, depending on which measurements are available:
\begin{enumerate}
\item Median of all Cepheid and/or tip of the red-giant branch distances: used for 10 galaxies.
\item Median of all distance estimates obtained with any of the following methods: planetary-nebulae luminosity function, brightest stars, radio and optical SNII, masers, globular-cluster luminosity function: used for 5 galaxies.
\item Median of all Tully-Fisher distances (including $d_\mathrm{T88}$): used for 8 galaxies.
\item \citet{tully_1988} estimate: used for 1 galaxy.
\end{enumerate}

It is worth noting that $d_\mathrm{best}$ is significantly different from $d_\mathrm{T88}$ ($>2$~Mpc) in about a third of all galaxies. For NGC~3198 and NGC~4414 $d_\mathrm{best}$ is obtained with method 1, and is respectively 3.7 and 8.1~Mpc larger than $d_\mathrm{T88}$. For NGC~4062, NGC~4274, UGC~2082, UGC~4278 and UGC~7774, $d_\mathrm{best}$ is obtained with method 3 and is always larger than $d_\mathrm{T88}$ (the last galaxy is the most extreme case, with a difference of 17.6 Mpc between the two estimates). In general, $d_\mathrm{T88}$ is always one of the lowest (when not the lowest) distance estimates available.

The method that we use to calculate star formation rates also bears some discussion. The galaxies that make up the HALOGAS sample are not otherwise included in a uniform observational sample at other wavelengths, in particular the infrared. We therefore use different methods for individual targets to calculate SFR, depending on what data are available. We prefer to use the combination of H$\alpha$ and TIR luminosity given by \citet{kennicutt_etal_2009}, where $L_{\mathrm{TIR}}$ is either calculated from IRAS or Spitzer fluxes via the prescriptions given by \citet{dale_helou_2002}. The IRAS form is used where possible, and the Spitzer form where IRAS fluxes are not available or provide only upper limits for the 25$\mu$m flux. IRAS fluxes are obtained from the references listed in Table \ref{table:sample}, and the Spitzer fluxes are from the Local Volume Legacy \citep[LVL;][]{dale_etal_2009}. H$\alpha$ fluxes are obtained first from \citet{kennicutt_etal_2008}, and for targets not included in that work, the literature was searched for fluxes. In all cases luminosities are calculated using the adopted distances listed in Table \ref{table:sample}. In two cases, UGC~2082 and NGC~4448, no H$\alpha$ fluxes are available; but since IRAS 60$\mu$m and 100$\mu$m fluxes are available, we calculate $L_{\mathrm{FIR}}$ using the equation given by \citet{rice_etal_1988}, and use SFR($L_{\mathrm{FIR}}$) as described by \citet{kennicutt_1998}. The references given in Table \ref{table:sample} were used to obtain the H$\alpha$ and IR fluxes. The derivations of the SFR for the HALOGAS sample will become more uniform with the release of the Wide-field Infrared Survey Explorer \citep[WISE;][]{wright_2008} data, which will also be well matched in angular resolution to our WSRT observations.

In Table \ref{table:environment}, we show environmental information about the target galaxies. The data are from the group catalog developed from the Nearby Optical Galaxies (NOG) sample \citep{giuricin_etal_2000}. The columns in the table are: (1) galaxy ID; (2) number of group members (using the percolation-1 grouping algorithm, see below); (3) NOG group (NOGG) number; (4) total group $B$ luminosity; (5) distance from group center of mass to outermost member; (6,7) minimum and maximum group member velocity relative to group center of mass; (8) contribution of the target to the group $B$ luminosity; (9) fraction of maximum group member $B$ luminosity; (10) normalized distance to group center of mass; (11) target velocity relative to group center of mass.

We have chosen to use group definitions resulting from the \citet{giuricin_etal_2000} ``P1'' algorithm (which is a percolation friends-of-friends algorithm that keeps the velocity grouping condition fixed irrespective of the galaxies' redshift). Because this definition of groups is based on angular and velocity coincidences, there will be spurious groupings or separations. We note corrections to the group definitions where appropriate in the text, and otherwise use the information as a means to illustrate the overall environmental characteristics of the HALOGAS sample. The survey galaxies sample a fairly broad range in environment. One quarter (6/24) of the HALOGAS sample reside in the same large group (Coma I). Three targets are members of pairs. In only one case is there a galaxy with a strongly interacting neighbor: NGC 672, which was included in the pilot survey described in this paper. The range of galaxy properties with respect to the group to which they belong also exhibits a wide range of values. The HALOGAS sample contains galaxies which dominate their group, as well as those which are minor group members and companions. If these smaller targets are gas rich, we may be probing gas accretion {\it donors} as well as recipients. Considering all of these properties, after observations of the full HALOGAS sample, we expect to be able to assess whether there is an environmental dependence on target galaxy halo properties.

\begin{table*}
\caption{Environmental properties of the HALOGAS sample galaxies.}
\label{table:environment}
\centering
\begin{tabular}{lcccccccccc}
\hline\hline
Galaxy & \# group & NOGG & $L_{B,\mathrm{group}}$ & $R_{\mathrm{group}}$ & $v_{\mathrm{min,group}}$ & $v_{\mathrm{max,group}}$ & $L_B/L_{B,\mathrm{group}}$ & $L_B/L_{B,\mathrm{max}}$ & $R/R_{\mathrm{group}}$ & $v$ \\
ID & members & number & [$10^{10}\,M_{\odot}$] & [Mpc] & [km/s] & [km/s] & $\,$ & $\,$ & $\,$ & [km/s] \\
\hline
NGC 0672              & 2   & 88               & 1.3  & 0.01 & -52  & 29  & 0.64 & 1.00 & 0.55 & 29 \\
NGC 0891$^\mathrm{a}$ & $-$ & $-$              & $-$  & $-$  & $-$  & $-$ & $-$  & $-$  & $-$  & $-$ \\   
NGC 0925              & 6   & 149              & 3.1  & 0.56 & -16  & 47  & 0.69 & 1.00 & 0.25 & -16 \\
NGC 0949              & 6   & 149              & 3.1  & 0.56 & -16  & 47  & 0.15 & 0.21 & 1.00 & 47 \\
UGC 2082$^\mathrm{b}$ & $-$ & $-$              & $-$  & $-$  & $-$  & $-$ & $-$  & $-$  & $-$  & $-$ \\   
NGC 1003$^\mathrm{a}$ & $-$ & $-$              & $-$  & $-$  & $-$  & $-$ & $-$  & $-$  & $-$  & $-$ \\
NGC 2403              & 2   & 319              & 1.8  & 0.24 & -14  & 1   & 0.88 & 1.00 & 0.13 & 1 \\
UGC 4278              & 6   & 345              & 1.4  & 0.77 & -90  & 35  & 0.30 & 1.00 & 0.44 & 21 \\
NGC 2541              & 6   & 345              & 1.4  & 0.77 & -90  & 35  & 0.24 & 0.79 & 0.21 & 35 \\
NGC 3198              & 2   & 462              & 1.7  & 0.43 & -25  & 1   & 0.96 & 1.00 & 0.05 & 1 \\
NGC 4062              & 27  & 631$^\mathrm{c}$ & 21.0 & 1.27 & -163 & 525 & 0.03 & 0.07 & 0.86 & 45 \\
NGC 4244              & 16  & 636              & 3.1  & 0.80 & -93  & 83  & 0.22 & 0.66 & 0.31 & -42 \\
NGC 4258              & 21  & 644              & 9.0  & 1.07 & -103 & 160 & 0.51 & 1.00 & 0.24 & -40 \\
NGC 4274              & 27  & 631$^\mathrm{c}$ & 21.0 & 1.27 & -163 & 525 & 0.04 & 0.10 & 0.45 & 194 \\
NGC 4414              & 27  & 631$^\mathrm{c}$ & 21.0 & 1.27 & -163 & 525 & 0.05 & 0.13 & 0.21 & -3 \\
NGC 4448              & 27  & 631$^\mathrm{c}$ & 21.0 & 1.27 & -163 & 525 & 0.02 & 0.06 & 0.37 & -79 \\
NGC 4559              & 27  & 631$^\mathrm{c}$ & 21.0 & 1.27 & -163 & 525 & 0.12 & 0.30 & 0.42 & 79 \\
NGC 4565              & 3   & 648              & 14.9 & 0.26 & -18  & 107 & 0.78 & 1.00 & 0.25 & -18 \\
UGC 7774              & 21  & 644              & 9.0  & 1.07 & -103 & 160 & 0.01 & 0.01 & 0.74 & 3 \\
NGC 4631              & 27  & 631$^\mathrm{c}$ & 21.0 & 1.27 & -163 & 525 & 0.40 & 1.00 & 0.30 & -91 \\
NGC 5023              & 7   & 723              & 8.0  & 0.45 & -94  & 131 & 0.04 & 0.09 & 0.68 & -77 \\
NGC 5055              & 7   & 723              & 8.0  & 0.45 & -94  & 131 & 0.37 & 0.79 & 1.00 & 8 \\
NGC 5229              & 7   & 723              & 8.0  & 0.45 & -94  & 131 & 0.01 & 0.02 & 0.97 & -94 \\
NGC 5585              & 8   & 766              & 3.1  & 0.54 & 107  & 75  & 0.07 & 0.09 & 0.52 & 75 \\
\hline
\end{tabular}
\begin{list}{}{}
\item[$^\mathrm{a}$] Not included in the \citet{giuricin_etal_2000} catalog due to low galactic latitude.
\item[$^\mathrm{b}$] No environmental information available; UGC~2082 is isolated.
\item[$^\mathrm{c}$] NOGG number 631 corresponds to Coma I.
\end{list}
\end{table*}

Because the HALOGAS galaxies are all nearby, there is the added advantage that most have been extensively observed at other wavelengths, for example as part of recent surveys such as the Spitzer Infrared Nearby Galaxies Survey \citep[SINGS;][]{kennicutt_etal_2003} and the 11~Mpc H$\alpha$ and Ultraviolet Galaxy Survey \citep[11HUGS;][]{lee_etal_2007}. The star formation process is clearly of interest with relation to the extraplanar, halo, and accreting gas that we are tracing, and we will therefore benefit by having these data available for our investigation. Although the available data will help to characterize disk star formation rates, we note that the coverage is not enough to cover the full WSRT primary beam. Furthermore, we are pursuing deep observations at other observational facilities to fill gaps where necessary (see \S\,\ref{section:multiwavelength}).

\section{Observations and data reduction}\label{section:observations}

The data are being obtained using the WSRT, which is an east-west interferometer consisting of 10 fixed (of which one is not used) and 4 movable 25-m antennas. Baselines range from 36~m to 2.7~km, and the fixed antennas are on a regular grid with a spacing of 144~m. We used the array in its Maxishort configuration, which is designed to optimize the imaging performance for extended sources in individual tracks. The correlator was set up to provide two linear polarizations in 1024 channels over a 10 MHz bandwidth centered at the systemic velocity of each target. Each galaxy was observed for 10 full 12-hour tracks, and each track was bracketed by standard flux calibrator sources.

Data reduction was performed using usual techniques in the Miriad software package \citep{sault_etal_1995}. A standard data reduction path was used for most targets; exceptions were necessary due to individual circumstances and will be described where appropriate. First, the raw data were visually inspected, and samples corrupted by radio frequency interference (RFI) were flagged where necessary. System temperature variations, which are measured and recorded at each antenna during the observations, were corrected for in both the calibrator and target data sets. Next, the standard calibrator sources were used to set the flux scale, as well as define and correct for the bandpass response. The radio continuum emission in the field was subtracted in the visibility domain and saved as ``channel-0'' data. An imaging and self-calibration loop was used on the continuum data to solve for time-variable antenna gain phases. The resulting antenna gains were subsequently applied to the continuum-subtracted spectral line data. The \HI\ data were imaged with a variety of weighting schemes (described below), creating several different data cubes for each target (see Table \ref{table:cubes}). Offline Hanning smoothing led to a final velocity resolution of $4.12\,\mathrm{km\,s^{-1}}$, although we may choose to degrade this resolution in post-processing in order to maximize sensitivity to features with broad velocity width. Finally, Clark CLEAN deconvolution was performed in two stages, the first within masked regions containing the brightest line emission, and the second within the entire imaged region.

Different weighting schemes are used in inverting the data from the $u,v$ plane into the image domain. In practice, within miriad's {\tt invert} task, we use a robust parameter of $0$ for intermediate resolution and sensitivity. To maximize sensitivity to faint extended emission we additionally use a Gaussian $u,v$ taper corresponding to $30\arcsec$ in the image plane. See Figures \ref{figure:u2082}, \ref{figure:n0672}, \ref{figure:n0925}, and \ref{figure:n4565}.

For the pilot survey galaxies presented in this paper, we list in Table \ref{table:cubes} a summary of the data cubes which have been produced and analyzed for the purposes of the presentation in \S\,\ref{section:results}. The columns are (1) Galaxy ID; (2) cube description; (3,4) beam size in arcsec and in kpc (note, the beam position angle is in all cases very close to $0^\circ$ because WSRT is an east-west array); (5) rms noise level; (6) the $3\sigma$ column density sensitivity to emission with a velocity width of $16\,\mathrm{km\,s^{-1}}$; and (7) the \HI\ mass for a $3\sigma$ detection of a spatially unresolved source with the same $16\,\mathrm{km\,s^{-1}}$ velocity width. The \HI\ masses are calculated using $d_\mathrm{best}$ from Table \ref{table:sample}.

\begin{table*}
\caption{Properties of Pilot Survey Datacubes}
\label{table:cubes}
\centering
\begin{tabular}{llcccccc}
\hline\hline
Galaxy & Visibility & Beam size & Beam size & rms noise & $3\sigma$ column density & $3\sigma$ \HI\ mass \\
ID & weighting & (arcsec) & (kpc) & ($\mathrm{mJy\,beam}^{-1}$) & sensitivity ($10^{19}\,\mathrm{cm}^{-2}$) & sensitivity ($10^5\,M_\odot$) \\
\hline
UGC~2082 & uniform             & $24.4\times11.3$ & $1.7\times0.8$ & 0.28 & 5.4 & 6.6 \\
\,       & robust-0            & $29.1\times12.6$ & $2.0\times0.9$ & 0.21 & 3.0 & 4.9 \\
\,       & $30\arcsec$ taper   & $40.9\times33.1$ & $2.9\times2.3$ & 0.20 & 0.8 & 4.7 \\
NGC~672  & uniform             & $27.4\times12.3$ & $1.0\times0.5$ & 0.28 & 4.4 & 1.8 \\
\,       & robust-0            & $30.6\times13.5$ & $1.1\times0.5$ & 0.21 & 2.7 & 1.4 \\
\,       & $30\arcsec$ taper   & $41.7\times32.0$ & $1.5\times1.2$ & 0.19 & 0.8 & 1.2 \\
NGC~925  & uniform             & $22.2\times12.6$ & $1.0\times0.6$ & 0.24 & 4.5 & 2.3 \\
\,       & robust-0            & $25.3\times13.9$ & $1.1\times0.6$ & 0.18 & 2.7 & 1.7 \\
\,       & $30\arcsec$ taper   & $37.9\times33.2$ & $1.7\times1.5$ & 0.17 & 0.7 & 1.6 \\
NGC~4565 & uniform             & $27.4\times12.0$ & $1.4\times0.6$ & 0.24 & 3.9 & 3.2 \\
\,       & robust-0            & $31.9\times13.8$ & $1.7\times0.7$ & 0.17 & 2.0 & 2.3 \\
\,       & $30\arcsec$ taper   & $43.7\times33.7$ & $2.3\times1.8$ & 0.23 & 0.8 & 3.0 \\
\hline
\end{tabular}
\end{table*}

The data cubes were used to generate images of the total \HI\ line intensity (moment-0 images), and of the intensity-averaged line-of-sight velocity (first-moment images). These were produced by smoothing the data cubes to a resolution of $90\arcsec$, clipping the original cubes at the $3\sigma$ level of the smoothed cubes, and performing the moment calculations mentioned above. The miriad task {\tt moment} was used for this process. In the case of the total intensity map, the primary beam correction was applied (for the WSRT 25-m antennas, the primary beam shape takes the mathematical form $\cos^6(cr\nu)$, where $c$ is a constant with a value of 68 at L-band, $\nu$ is the frequency in GHz, and $r$ is the field radius in radians). Then, intensities were converted to column densities ($N_{\mathrm{HI}}$) assuming optically thin gas. The resulting maps are displayed and discussed in \S\,\ref{section:results}. We note that typical velocity profiles are more complicated than can be encapsulated in a single number such as the average velocity. Especially in edge-on galaxies, first-moment maps can be misleading, and therefore for our two highly inclined targets we instead show Renzograms (a single outer contour for each of a series of velocity channels, plotted on the same map).

The primary goal of the survey is to characterize each galaxy in the sample by the properties of the extraplanar gas (if any) detected in the field, and to consider the full sample together to draw conclusions about the global characteristics of this phenomenon. A detailed picture of the individual galaxies will be one of the output products of this investigation. The results for such individual survey targets will be presented in a series of companion papers. The main engine for analyzing these galaxies will be tilted-ring modeling software, which builds geometrical galaxy models using a series of idealized, roughly concentric, tilted rings. The tilted-ring modeling software is based in the GIPSY package. We will use a combination of GIPSY's GALMOD, and the Tilted Ring Fitting Code (TiRiFiC) which has been developed by \citet{jozsa_etal_2007}.

\section{Overview of Pilot Survey Observations}\label{section:results}

In this Section, we present an overview of the data obtained during our Pilot Survey, which is made up of four targets observed during the first observing semester of the HALOGAS survey. The targets presented here are representative of a broad range of galaxy characteristics, and give a sense of the diversity in the full HALOGAS sample. For each galaxy, we produced a continuum map created using all $10\times12$hr of data, resulting in rms noise levels of about $30\,\mu\mathrm{Jy\,beam^{-1}}$. All targets are detected in continuum emission. We present maps of the total \HI\ line intensity, comparisons with the optical image, and the \HI\ first-moment image or Renzogram.

For each galaxy we provide some discussion of the target and the results gleaned from our HALOGAS Pilot Survey data products. In the interest of brevity, we defer fuller detailed analyses of these and certain other individual HALOGAS targets (including NGC~4244; Zschaechner et al., in prep., UGC~4278; Serra et al., in prep., and NGC~949; J{\'o}zsa et al., in prep.) to forthcoming papers. Ultimately, the main goal of the HALOGAS project is to examine the properties of extraplanar \HI\ across the full sample; those results will be presented following the completion of all survey observations.

\subsection{UGC 2082}\label{subsection:u2082}

UGC~2082 is an edge-on, low surface brightness galaxy which appears to be isolated (hence the lack of environmental information in the UGC 2082 entry in Table \ref{table:environment}). It contains molecular gas \citep{matthews_gao_2001,matthews_etal_2005}, but has a low star formation rate. The SFR estimate listed in Table \ref{table:sample} ($0.023\,M_{\odot}\,\mathrm{yr}^{-1}$) is based on FIR luminosity, and by this estimator UGC~2082 has one of the lowest SFRs in our sample. We note, however, a confident detection of extended radio continuum emission from the inner disk.

With the HALOGAS \HI\ observations, we detect a total \HI\ mass of $M_{\mathrm{HI}}\,=\,2.54\times10^{9}\,M_\odot$.
The vertical distribution of \HI\ emission is somewhat extended, reaching up to a maximum vertical extent of about 5 kpc, fully one third of the radial extent of the \HI.
The total \HI\ contours along the major axis clearly show a slight warp in the sky plane, beginning somewhat beyond the optical radius. A warp component along the line of sight may be responsible for some of the apparent thickening of the disk.

\begin{figure*}
\resizebox{\hsize}{!}{\includegraphics{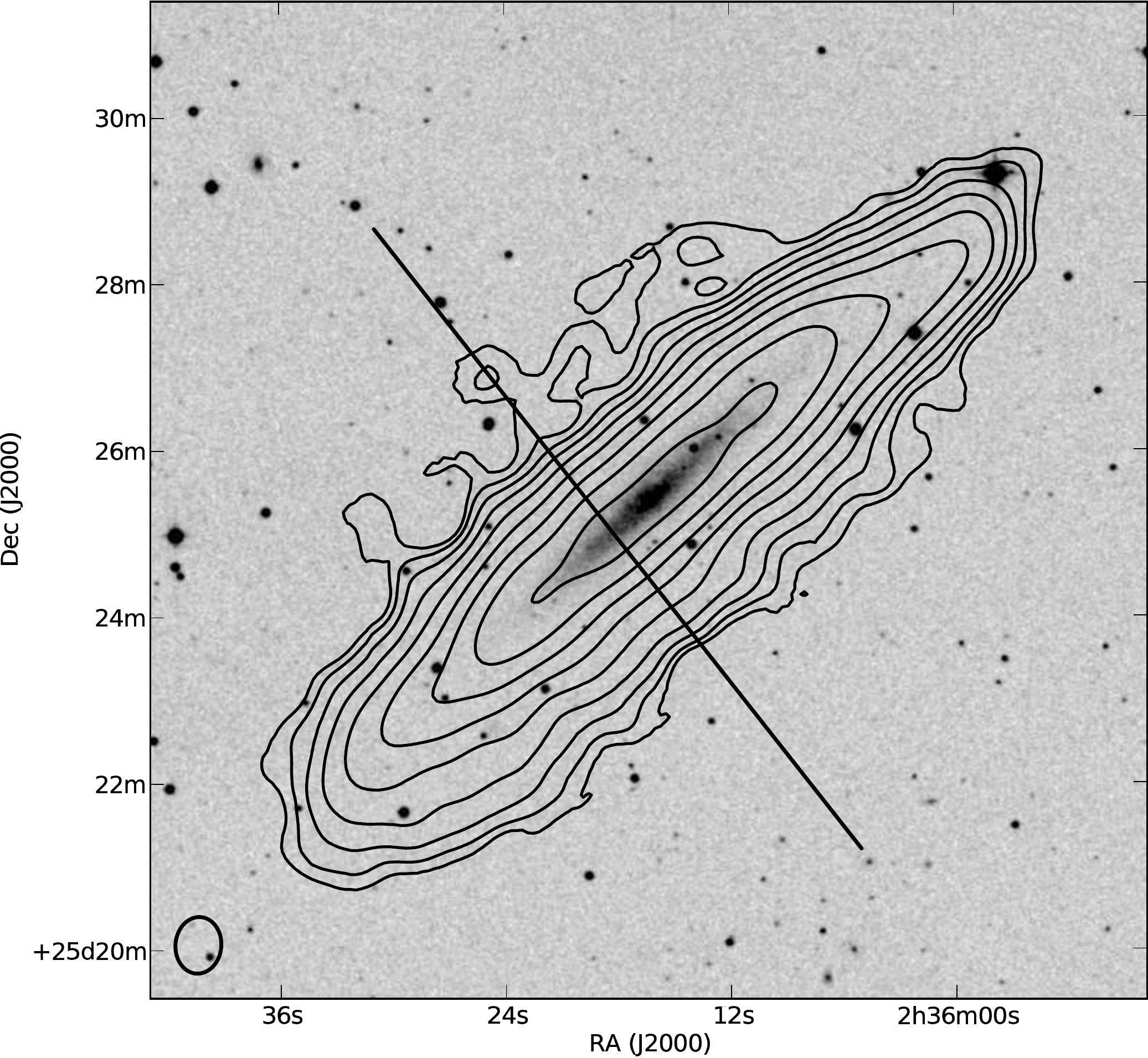}\includegraphics{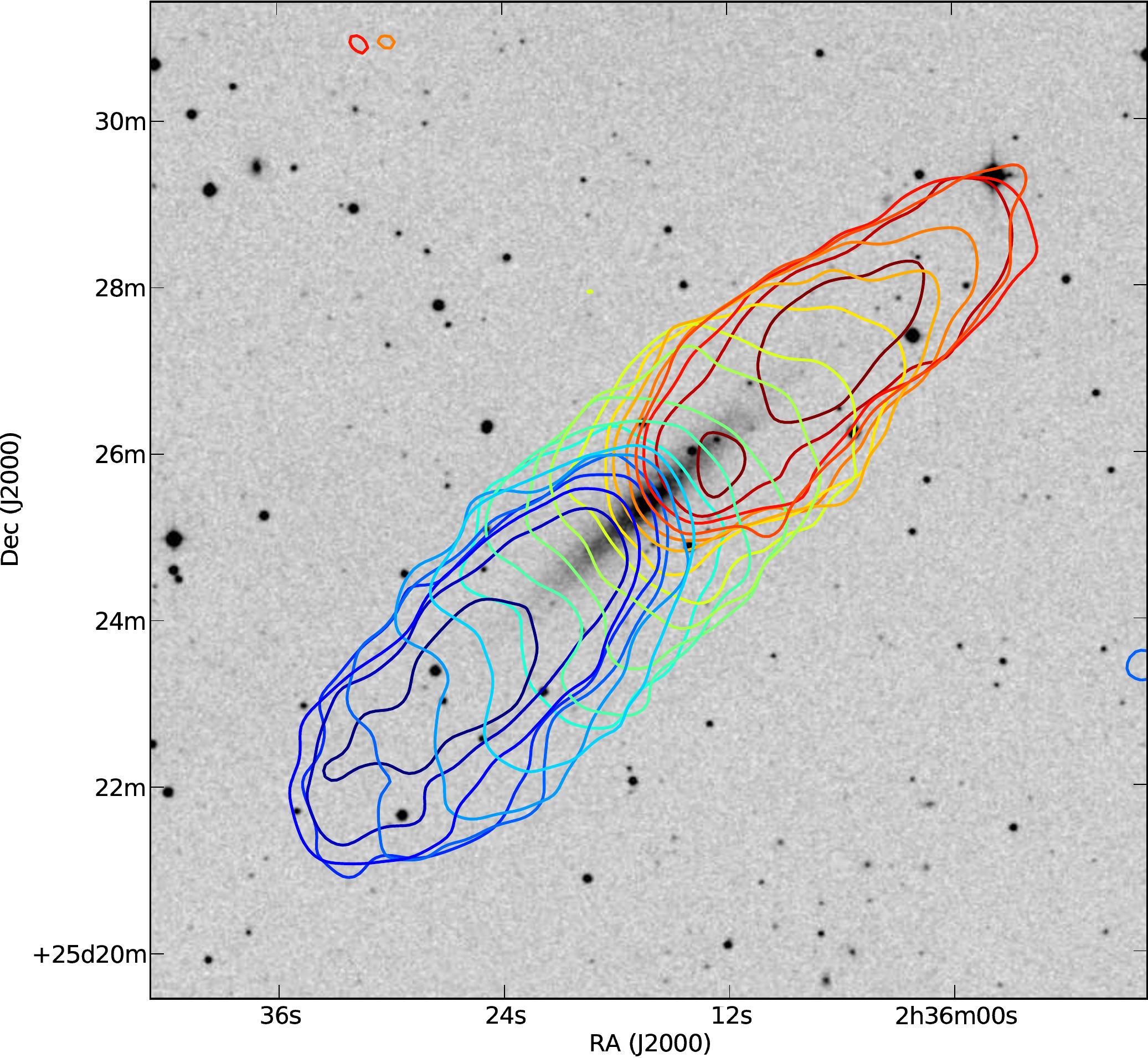}}
\caption{Overview of the HALOGAS observations of UGC~2082. The left panel shows the \HI\ total intensity overlaid on the DSS $R$-band image. The \HI\ contours originate from the $30\arcsec$-tapered image, begin at $N_{HI}\,=\,1.0\times10^{19}\,\mathrm{cm^{-2}}$ and increase by powers of two. The straight line shows the orientation of the PV slice shown in Figure \ref{figure:pvdiagram_u2082}. The right panel shows an overlay of several channels in the lowest resolution data cube, all at a level of $0.9\,\mathrm{mJy\,beam^{-1}}$ ($\approx3.75\sigma$). The contours are separated by $12.4\,\mathrm{km\,s^{-1}}$, begin at $593\,\mathrm{km\,s^{-1}}$ (dark blue) and range upward to $815\,\mathrm{km\,s^{-1}}$ (dark red). Both panels show the same area of the sky. The beam size of the \HI\ data is shown in the lower left corners of the left panel.}
\label{figure:u2082}
\end{figure*}

\begin{figure}
\resizebox{\hsize}{!}{\includegraphics{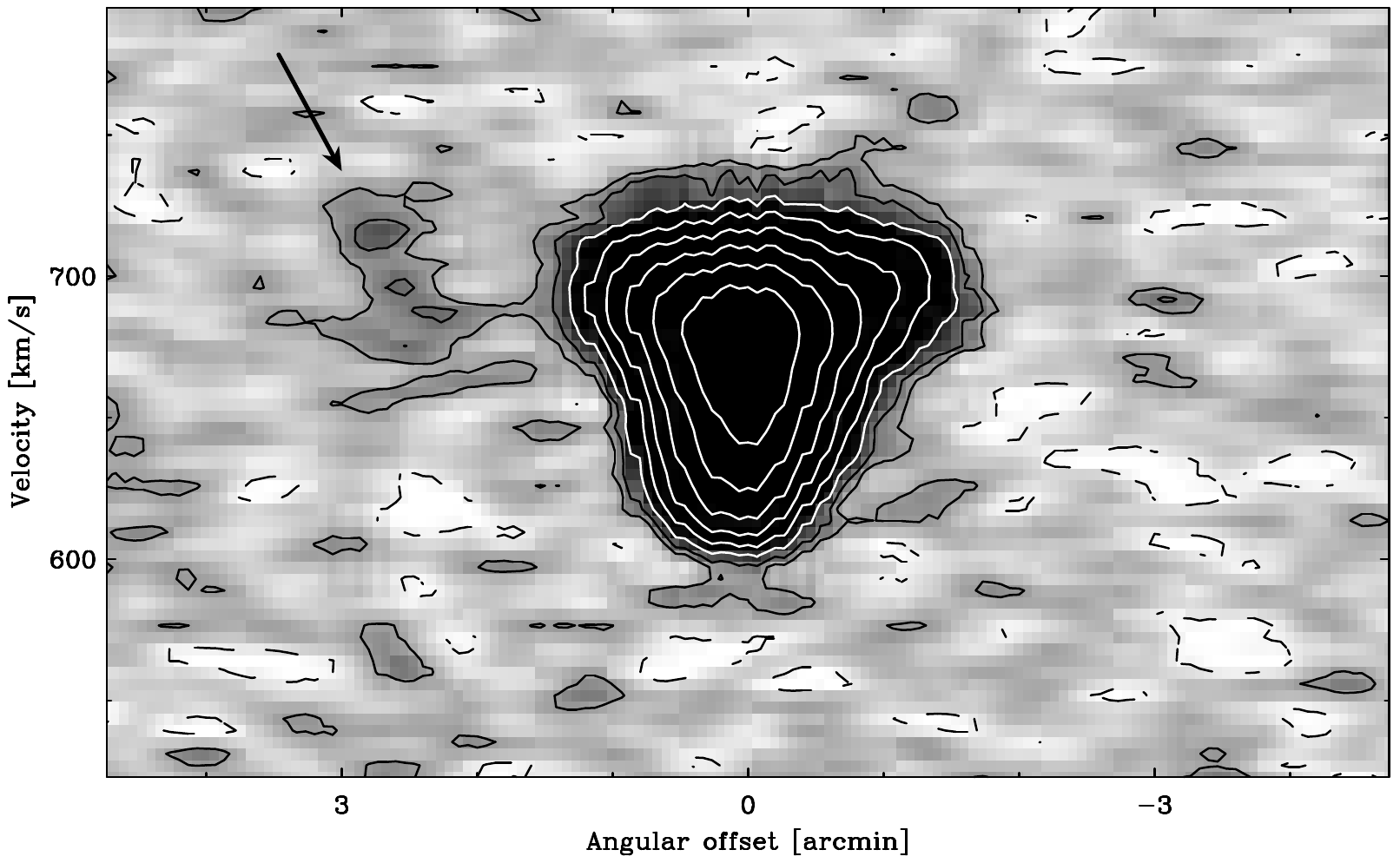}}
\caption{Position-velocity (PV) diagram for UGC~2082, plotted using the $30\arcsec$-resolution cube. Contour levels begin at $300\,\mu\mathrm{Jy\,beam^{-1}}$ ($\approx1.5\sigma$; see Table \ref{table:cubes}) and increase by powers of two. The dashed contours are at $-300\,\mu\mathrm{Jy\,beam^{-1}}$. The line in Figure \ref{figure:u2082} indicates the orientation of the PV slice shown here. The northeast end of the slice is to the left. At the distance listed in Table \ref{table:sample}, $1^\prime$ corresponds to a linear scale of $3.11\,\mathrm{kpc}$. The arrow indicates an extraplanar feature described in the text.}
\label{figure:pvdiagram_u2082}
\end{figure}

Indeed, the appearance of the position-velocity (PV) diagram shown in Figure \ref{figure:pvdiagram_u2082} is suggestive of a line-of-sight warp. In this diagram, the major axis is located at zero angular offset. The characteristic triangular appearance, caused by emission at large vertical offsets appearing at velocities close to systemic, can be caused by a halo with lagging rotational velocities or by a line-of-sight warp \citep[e.g.][]{oosterloo_etal_2007}. In the case of UGC~2082, the visual appearance (in particular the narrowness of the velocity profiles at the largest angular offsets) is more similar to galaxies determined to show evidence for a line-of-sight warp \citep[e.g.,][]{rand_benjamin_2008,gentile_etal_2003} rather than those determined to host a lagging halo. The Renzogram shown in Figure \ref{figure:u2082} also shows characteristic signs of a line-of-sight warp: the contours of channels at intermediate velocities extend to larger radii with increasing distance from the midplane. That this effect is more pronounced on the approaching side, together with the uneven thickness of the \HI\ distribution along the major axis, indicate that the line-of-sight warp is asymmetric. Geometrical modeling work is in progress to draw conclusions about how much of this vertical extension can be due to a halo, rather than warping (or flaring). Based on modeling as shown by \citet[][their Figure 4]{swaters_etal_1997} and \citet[][their Figure 12]{oosterloo_etal_2007}, the different situations can be distinguished by careful inspection of the individual channel maps, as well as PV diagrams such as the one shown in Figure \ref{figure:pvdiagram_u2082}. A detailed description of the modeling work is beyond the scope of this pilot survey description paper, and it will instead be presented in a forthcoming publication.

It is unlikely that large-scale morphological features such as warping or flaring can explain all of the high-latitude \HI\ emission, however. The ``fingers'' (filamentary features) of \HI, extending upward perpendicular to the major axis at the largest $z$ on the northeast side, appear to be distinct enhancements of cold gas. One such feature is indicated with an arrow in Figure \ref{figure:pvdiagram_u2082}. At a projected distance of $\approx9\,\mathrm{kpc}$ above the midplane, this cloud would have an \HI\ mass of a few $10^6\,M_\odot$ and a broad velocity width of $60\,\mathrm{km\,s^{-1}}$. The presence of features like this one suggests that at least some of the thick disk of UGC~2082 is made up of a filamentary \HI\ halo.

\subsection{NGC 672}\label{subsection:n0672}

\begin{figure*}
\resizebox{\hsize}{!}{\includegraphics{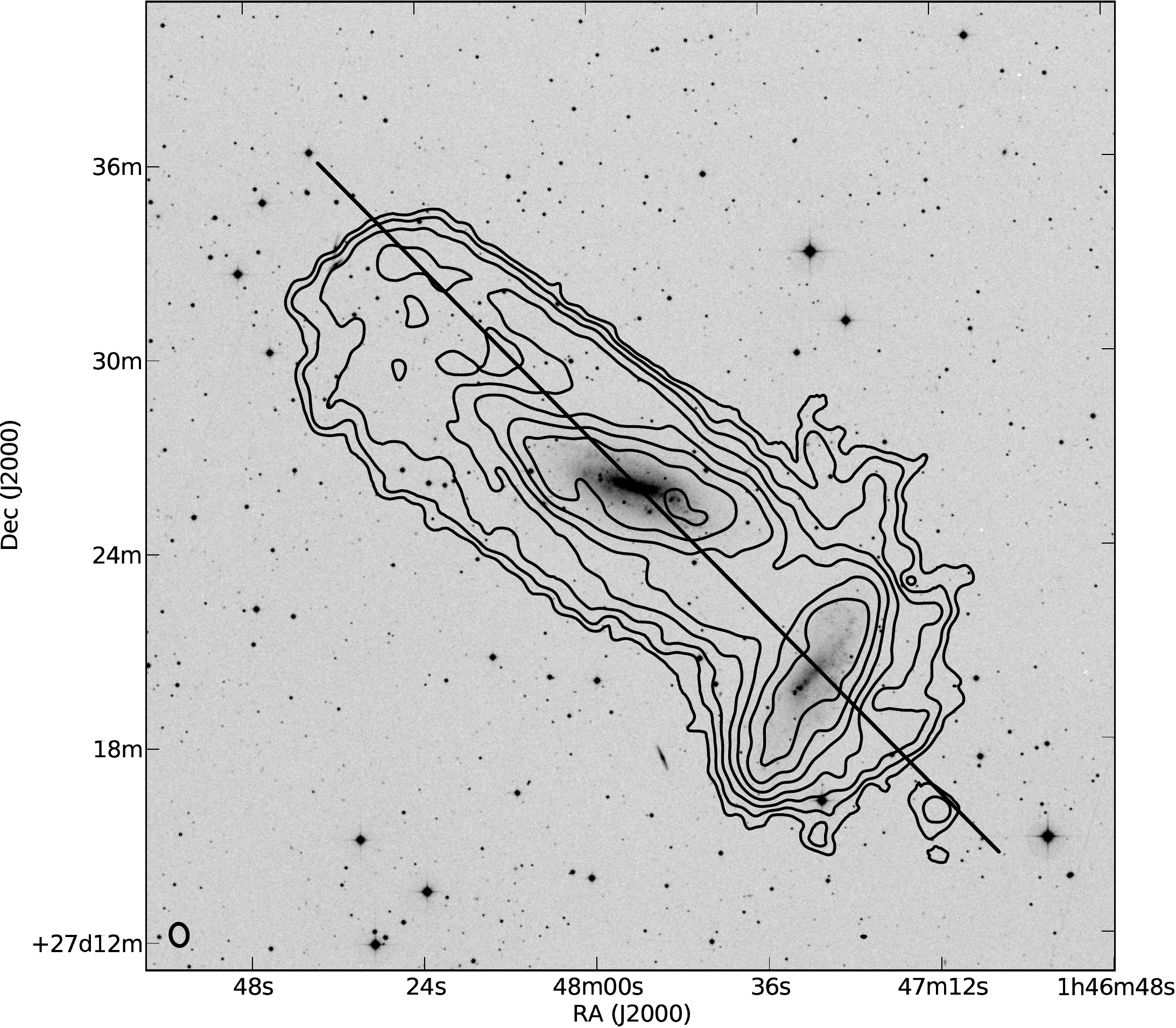}\includegraphics{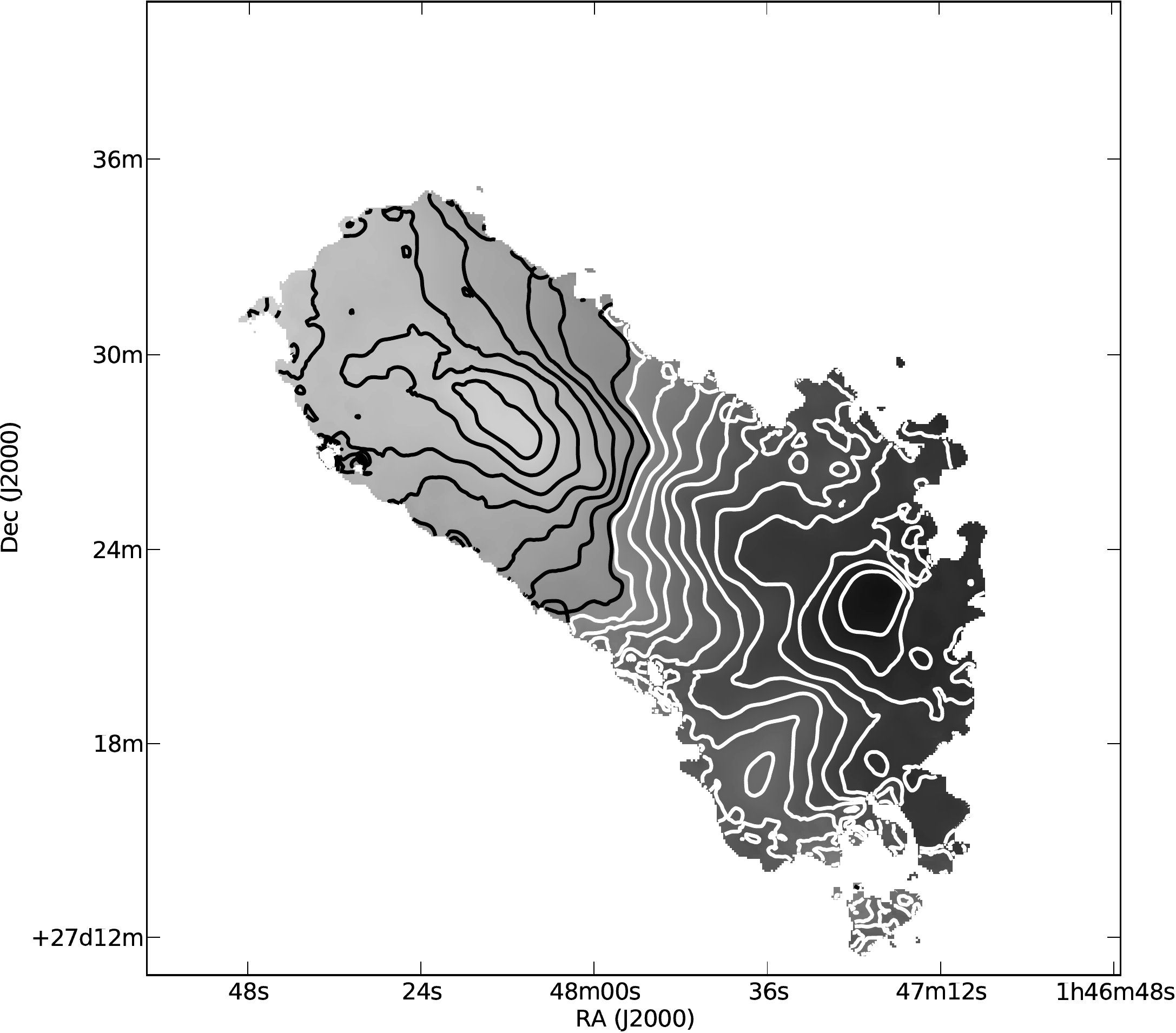}}
\caption{Overview of the HALOGAS observations of NGC~672. The left panel is as in Figure \ref{figure:u2082}. The contours begin at $N_{HI}\,=\,2.5\times10^{19}\,\mathrm{cm^{-2}}$ and increase by powers of two. The right panel shows the first-moment image derived from the lowest-resolution \HI\ cube. The grayscale ranges from $245\,\mathrm{km\,s^{-1}}$ (black) to $570\,\mathrm{km\,s^{-1}}$ (white). Contours are white on the approaching side and black on the receding side, and run from $255\,\mathrm{km\,s^{-1}}$ to $600\,\mathrm{km\,s^{-1}}$ in steps of $15\,\mathrm{km\,s^{-1}}$. The white contour closest to the center of the primary galaxy, corresponding to $425\,\mathrm{km\,s^{-1}}$, represents the systemic velocity of NGC~672.}
\label{figure:n0672}
\end{figure*}

NGC~672 is a moderately inclined spiral which has a neighboring irregular galaxy, IC~1727. In Figure \ref{figure:n0672}, NGC~672 is the brighter galaxy centered in the frames, and IC~1727 is located to its southwest. At the adopted distance listed in Table \ref{table:sample} (7.6~Mpc), the linear projected separation between the two galaxies is 16.3~kpc. In the optical image, there is no clear sign of a strong disturbance in NGC~672. Previous interferometric observations \citep{combes_etal_1980} have revealed a gravitational interaction with a substantial amount of gas in the interface region. In our observations, we detect a total mass of $3.78\times10^9\,M_{\odot}$, of which approximately 60\% is associated with the primary, NGC~672. Both galaxies are clearly detected in continuum emission.

The \HI\ distribution shown in Figure \ref{figure:n0672} reveals clearly the connection between the two galaxies. Each of the two galaxies appears to be relatively undisturbed within their optical extent, but the outskirts and interface region are more complicated. The pair seems to be in an early stage of interaction. A clear enhancement of \HI\ emission is located between the two galaxies, centered at about $\alpha_{\mathrm{J2000.0}}=01^{h}47^{m}41^{s}$, $\delta_{\mathrm{J2000.0}}=27^{d}22^{m}10^{s}$. To the northwest of the interaction region, a clear arc-shaped feature is seen extending away from the system. A faint diffuse extension is seen on the southwest side of IC~1727. A long, coherent arm-shaped feature starts just north of the center of NGC~672 and stretches away to the northeast. This could be a tidal or spiral arm, or may be a large \HI\ loop seen in projection.

\begin{figure}
\resizebox{\hsize}{!}{\includegraphics{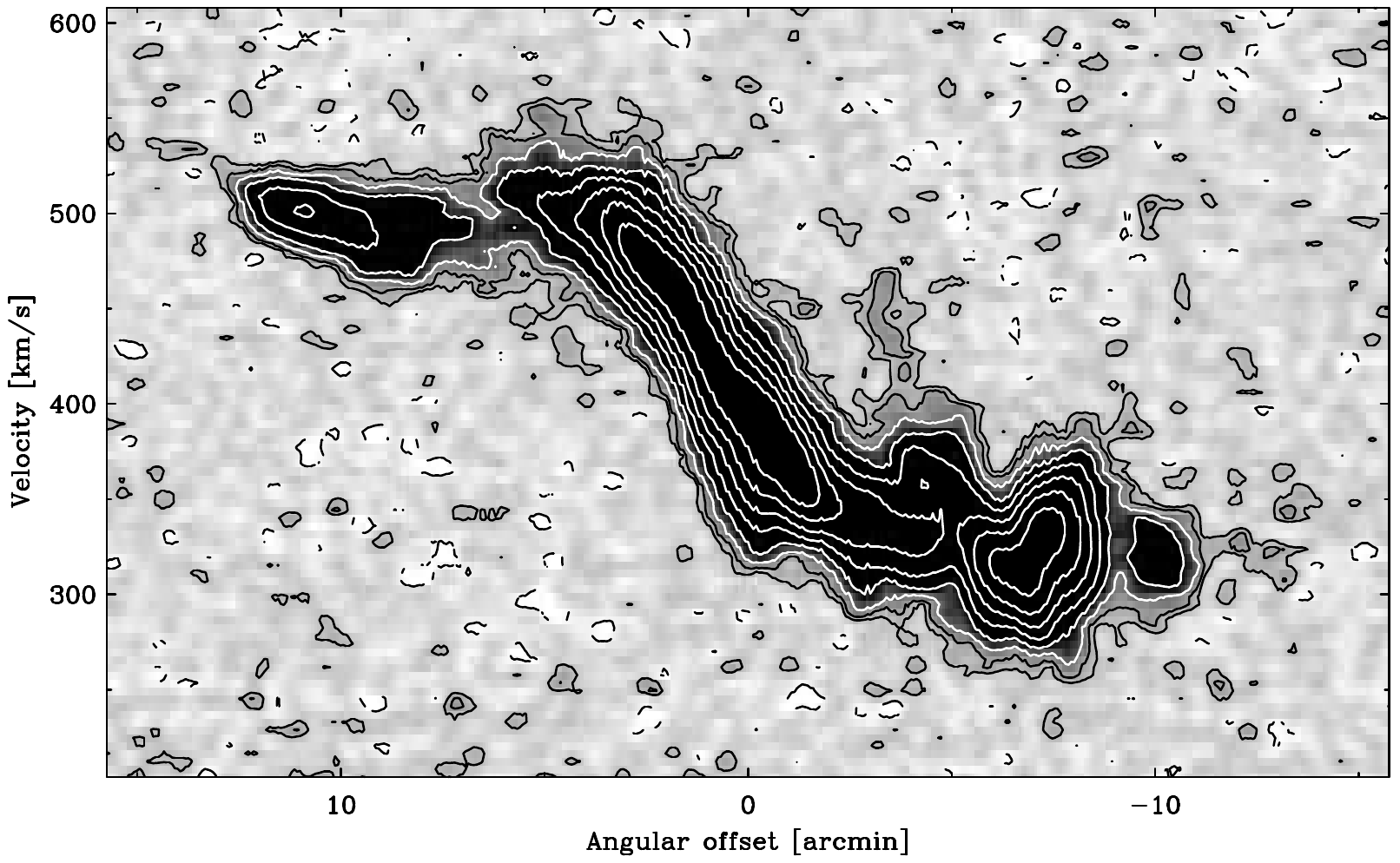}}
\caption{As in Figure \ref{figure:pvdiagram_u2082}, but for NGC~672. The line in Figure \ref{figure:n0672} indicates the orientation of the PV slice shown here. The northeast end of the slice is to the left. At the distance listed in Table \ref{table:sample}, $1^\prime$ corresponds to a linear scale of $2.18\,\mathrm{kpc}$.}
\label{figure:pvdiagram_n0672}
\end{figure}

The first-moment image indicates that the rotation of NGC~672 is fairly well ordered within the optical body on the northeast side (farther from the interaction region), and significantly more disturbed toward the companion galaxy. The main body of IC~1727 appears to show well ordered rotation. The gas in the system that is located outside of the optical disks also has kinematics which overall blend smoothly into the first-moment images of the individual galaxies. It should be noted, however, that the velocity profiles are complicated in this region; this effect is clearly visible in the PV diagram shown in Figure \ref{figure:pvdiagram_n0672} (the emission near angular offsets of $3\arcmin-4\arcmin$). There, the profiles are significantly broadened, often have multiple peaks, and even show evidence for discrete emission at distinctly different velocities (e.g. the feature at angular offset $\sim3\arcmin$, and velocity $420-460\,\mathrm{km\,s}^{-1}$). For that reason, the mean velocities in the interface region shown in Figure \ref{figure:n0672} should be regarded with suspicion. The PV diagram also shows that the systemic velocity of IC~1727 is consistent with a continuation of the kinematics of NGC~672. As noted by \citet{sancisi_etal_2008}, in many minor mergers the kinematics of the infalling companion galaxy follows that of the outer disk of the major galaxy.

In order to gain a better understanding of this system, we intend to use Identikit \citep{barnes_hibbard_2009}, or similar software, to match the observations against computer models of galaxy interactions. This should give an indication of the three-dimensional geometry of the system, as well as the interaction history. There is clearly a large quantity of unsettled gas in the system, which is potentially material that will be accreted at a later stage. The ultimate fate of this gas is of obvious interest.

\subsection{NGC 925}\label{subsection:n0925}

\begin{figure*}
\resizebox{\hsize}{!}{\includegraphics{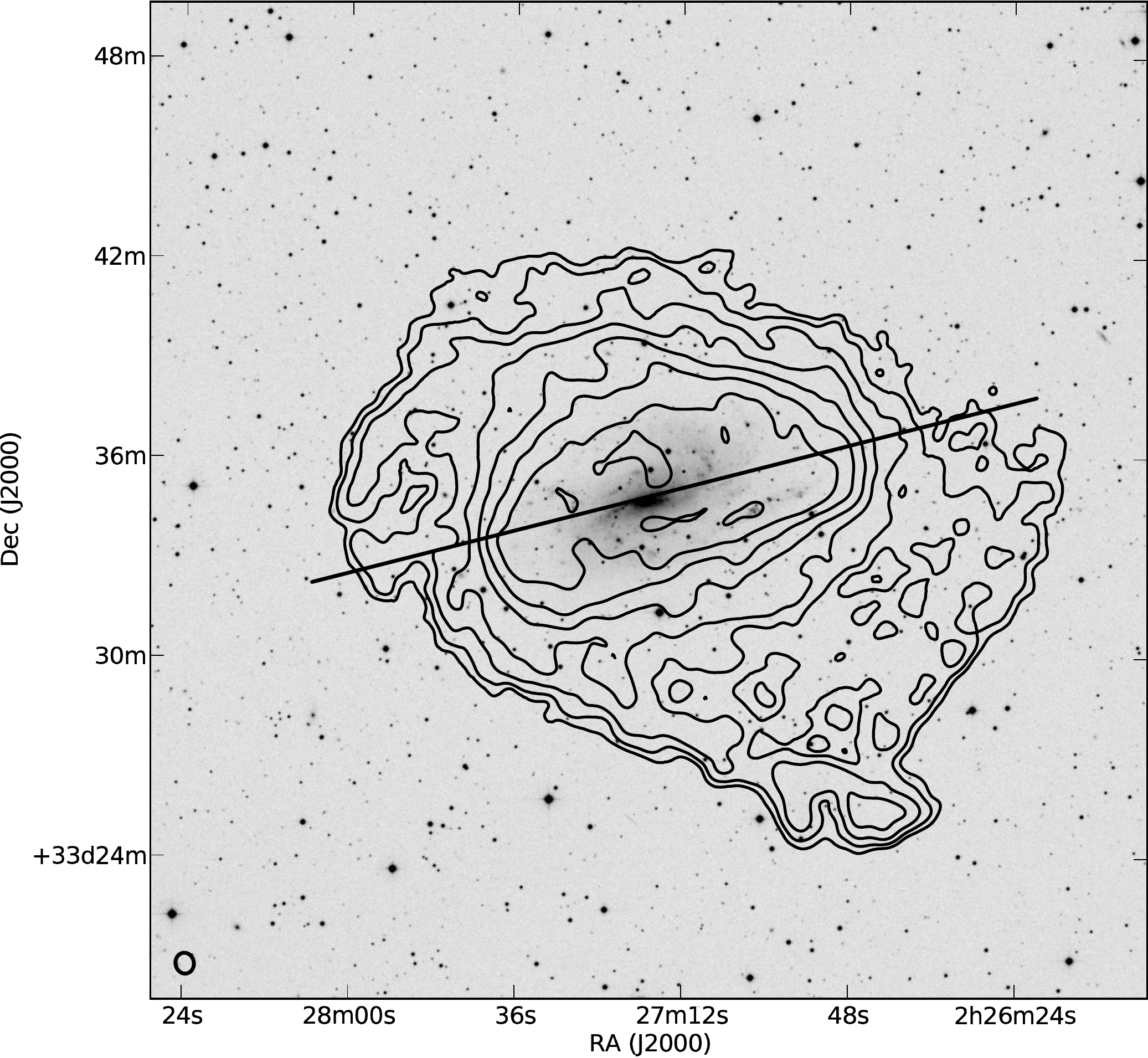}\includegraphics{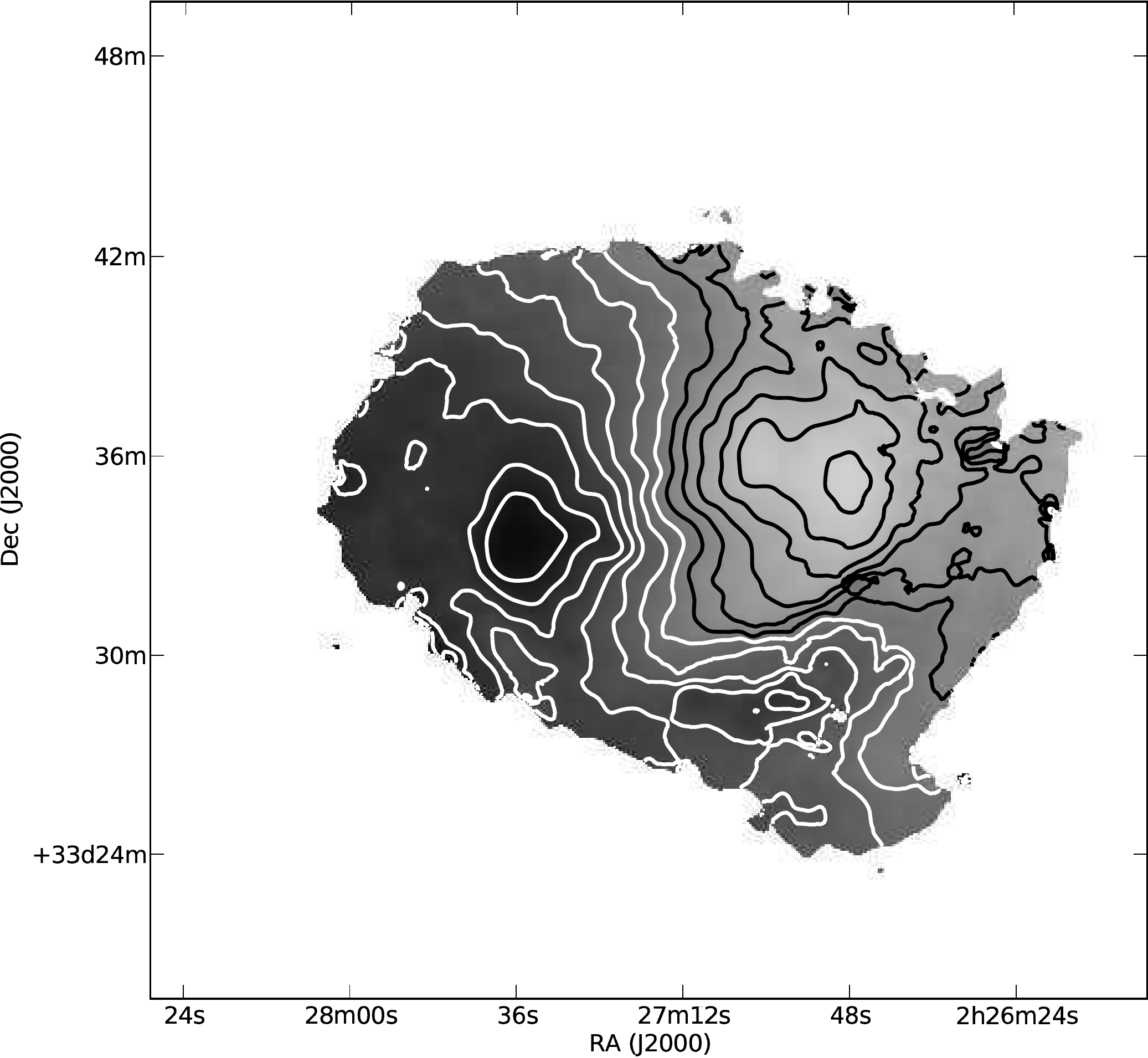}}
\caption{Overview of the HALOGAS observations of NGC~925. The panels are as in Figure \ref{figure:n0672}. The contours in the left panel begin at $N_{HI}\,=\,2.0\times10^{19}\,\mathrm{cm^{-2}}$ and increase by powers of two. The grayscale in the right panel ranges from $440\,\mathrm{km\,s^{-1}}$ (black) to $675\,\mathrm{km\,s^{-1}}$ (white). Contours are white on the approaching side and black on the receding side, and run from $450\,\mathrm{km\,s^{-1}}$ to $660\,\mathrm{km\,s^{-1}}$ in steps of $15\,\mathrm{km\,s^{-1}}$. The white contour closest to the center of the galaxy, corresponding to $555\,\mathrm{km\,s^{-1}}$, represents the systemic velocity of NGC~925.}
\label{figure:n0925}
\end{figure*}

NGC~925 is a moderately inclined spiral. We detect bright diffuse continuum emission, with enhancements corresponding to the locations of the most prominent features in the optical spiral arms. NGC~925 has been observed previously in the \HI\ line, notably by \citet{pisano_etal_1998} and as part of THINGS \citep[The \HI\ Nearby Galaxy Survey;][]{walter_etal_2008}. Neither of those studies detected \HI\ at such large galactocentric radii as revealed in Figure \ref{figure:n0925}. Note that the image shown by \citet{sancisi_etal_2008}, based on earlier WSRT data, clearly illustrates the radially extended and disturbed \HI\ distribution. With the HALOGAS observations, we recover the same structure, but with higher sensitivity. The inner disk within the optical body has an unsmooth appearance, with clumps and holes. Many of the \HI\ overdensities correspond to the locations of bright knots of optical light in the spiral arms. The appearance of the outer disk of NGC~925 is that of a complicated stream of gas stripped during a minor merger. The twisting of the isovelocity contours in that region suggest outer disk warping or spiral structure. An enhancement in the \HI\ column density is seen at the southernmost edge of the outer disk, at coordinates $\alpha_{\mathrm{J2000.0}}=02^{h}26^{m}44^{s}$, $\delta_{\mathrm{J2000.0}}=33^{d}25^{m}20^{s}$. Inspection of the optical DSS plates reveals that a faint fuzzy optical counterpart is very close to the location of this \HI\ enhancement, centered at about $\alpha_{\mathrm{J2000.0}}=02^{h}26^{m}53^{s}$, $\delta_{\mathrm{J2000.0}}=33^{d}25^{m}40^{s}$. Thus, the \HI\ streams may be the result of the disruption of a gas-rich dwarf companion, whose stellar counterpart is still barely visible in the survey plates. NGC~925 is a prototypical minor merger, illustrating how gas is fed into large galaxies by the accretion of material from satellites.

\begin{figure}
\resizebox{\hsize}{!}{\includegraphics{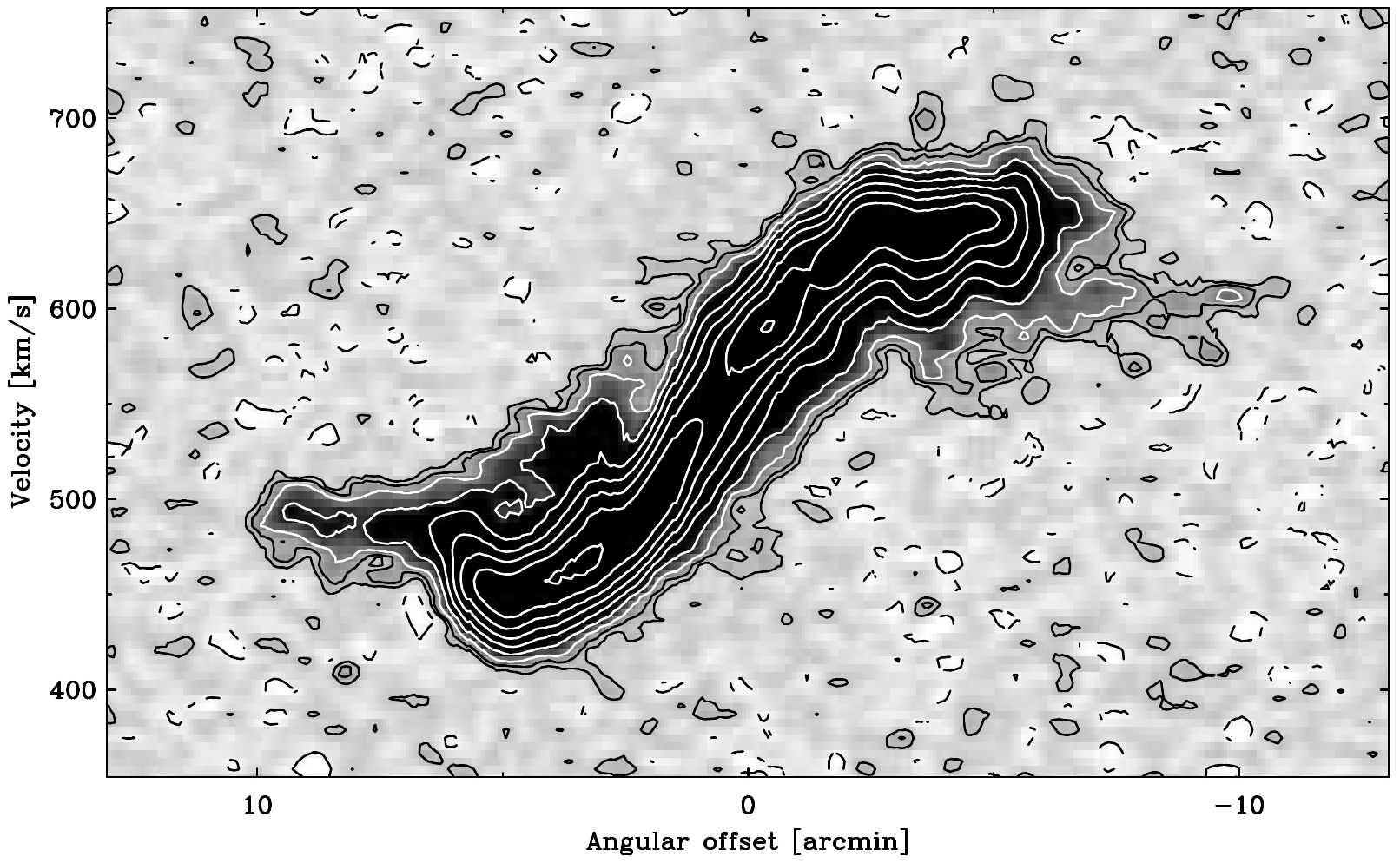}}
\caption{As in Figure \ref{figure:pvdiagram_u2082}, but for NGC~925. The line in Figure \ref{figure:n0925} indicates the orientation of the PV slice shown here. The east end of the slice is to the left. The lowest contour level (solid and dashed) is $\pm250\,\mu\mathrm{Jy\,beam^{-1}}$. At the distance listed in Table \ref{table:sample}, $1^\prime$ corresponds to a linear scale of $2.73\,\mathrm{kpc}$.}
\label{figure:pvdiagram_n0925}
\end{figure}

With the HALOGAS observations we detect $5.89\times10^9\,M_{\odot}$ of \HI. A large amount of anomalous \HI\ emission is clearly visible in this target. While the streams in the outer disk generally follow the overall rotation of the inner disk (see the first-moment image in Figure \ref{figure:n0925}), there is also gas in the inner parts of the galaxy which either appear as ``beard'' emission \citep{fraternali_etal_2002}, or sometimes even at distinct velocities relative to the bulk of the galaxy rotation. These features cannot be seen directly in the moment maps shown in Figure \ref{figure:n0925}. The beard gas is however readily visible in the PV diagram shown in Figure \ref{figure:pvdiagram_n0925}. There, faint emission is seen to extend preferentially toward the systemic velocity (to lower velocity at positive angular offset, and higher velocity at negative angular offset). This may well be the sign of a slowly rotating halo seen in projection against the disk. Interpreted in this way, we estimate the total \HI\ mass of the lagging component to be a few times $10^8\,M_\odot$, at a velocity difference of about $20-50\,\mathrm{km\,s^{-1}}$ with respect to the disk. We note that the SFR of NGC~925 is similar to that of NGC~2403, which is the prototypical galaxy demonstrating beard emission, and the latter is somewhat more massive. Detailed modeling of NGC~925 will allow us to separate the regularly rotating disk emission from the anomalous gas, determine its nature, and make a more detailed comparison with NGC~2403.

A small companion (outside of the frame in Figure \ref{figure:n0925}) is found to the north of NGC~925 at about $\alpha_{\mathrm{J2000.0}}=02^{h}27^{m}20^{s}$, $\delta_{\mathrm{J2000.0}}=33^{d}57^{m}30^{s}$. Its \HI\ mass is $1.27\times10^7\,M_{\odot}$. This object, too, appears to have a faint optical counterpart in the DSS plates. It is located at a projected distance of 60 kpc from the center of NGC~925 and does not show any obvious signs of interaction.

\subsection{NGC 4565}\label{subsection:n4565}

NGC~4565 is a large, edge-on spiral galaxy which is similar to the Milky Way. The new HALOGAS observations are combined with archival data presented by \citet{dahlem_etal_2005}. The other two galaxies in the frame are NGC~4562, a small companion classified in the RC3 as type SB(s)dm and visible to the southwest of NGC~4565, and IC~3571, the small galaxy located very close to NGC~4565 on its north side. In the PV diagram shown in Figure \ref{figure:pvdiagram_n4565}, the vertically extended emission in NGC~4565 is seen to have a smooth velocity gradient directly toward the systemic velocity of IC~3571.

\begin{figure*}
\resizebox{\hsize}{!}{\includegraphics{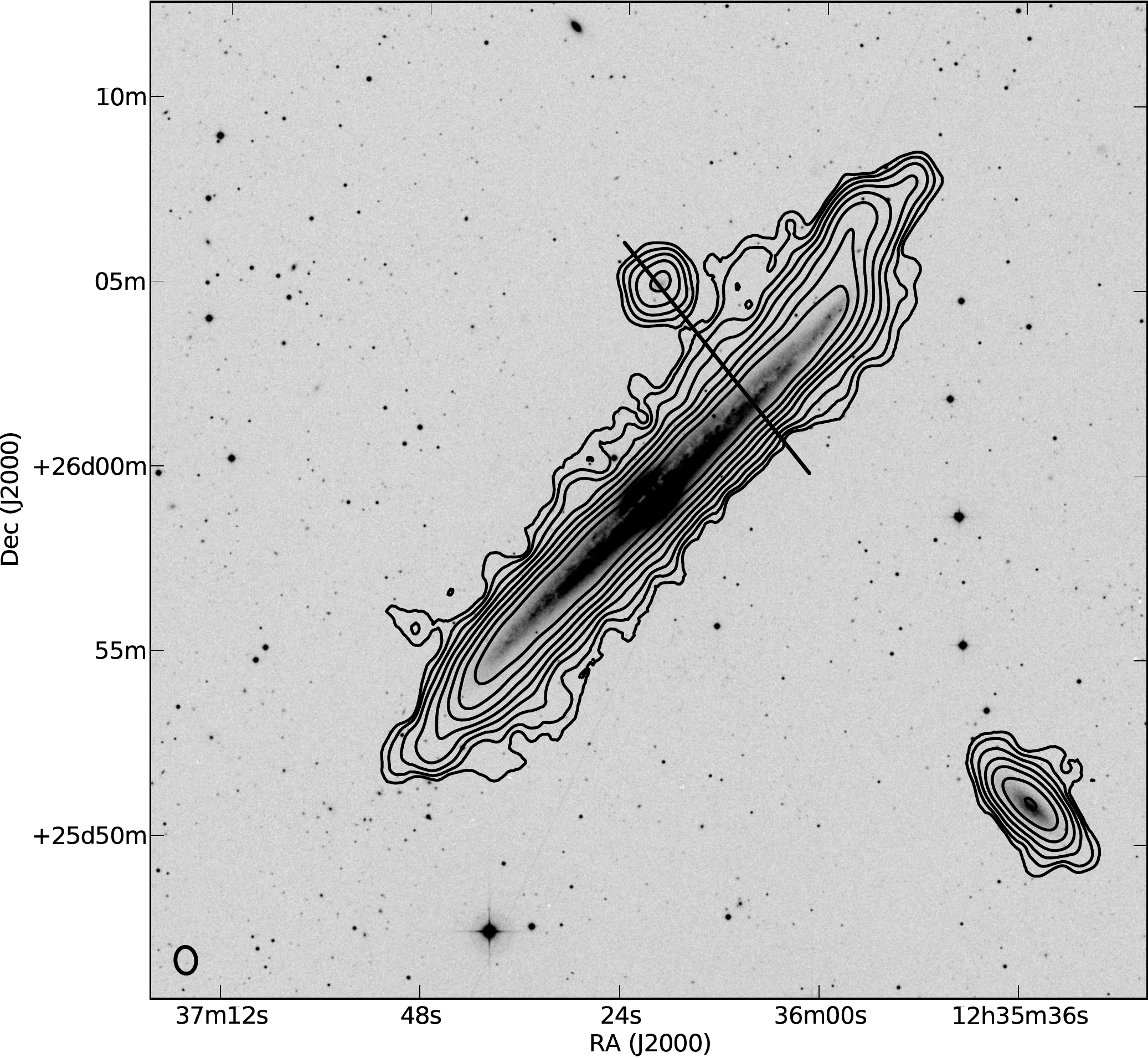}\includegraphics{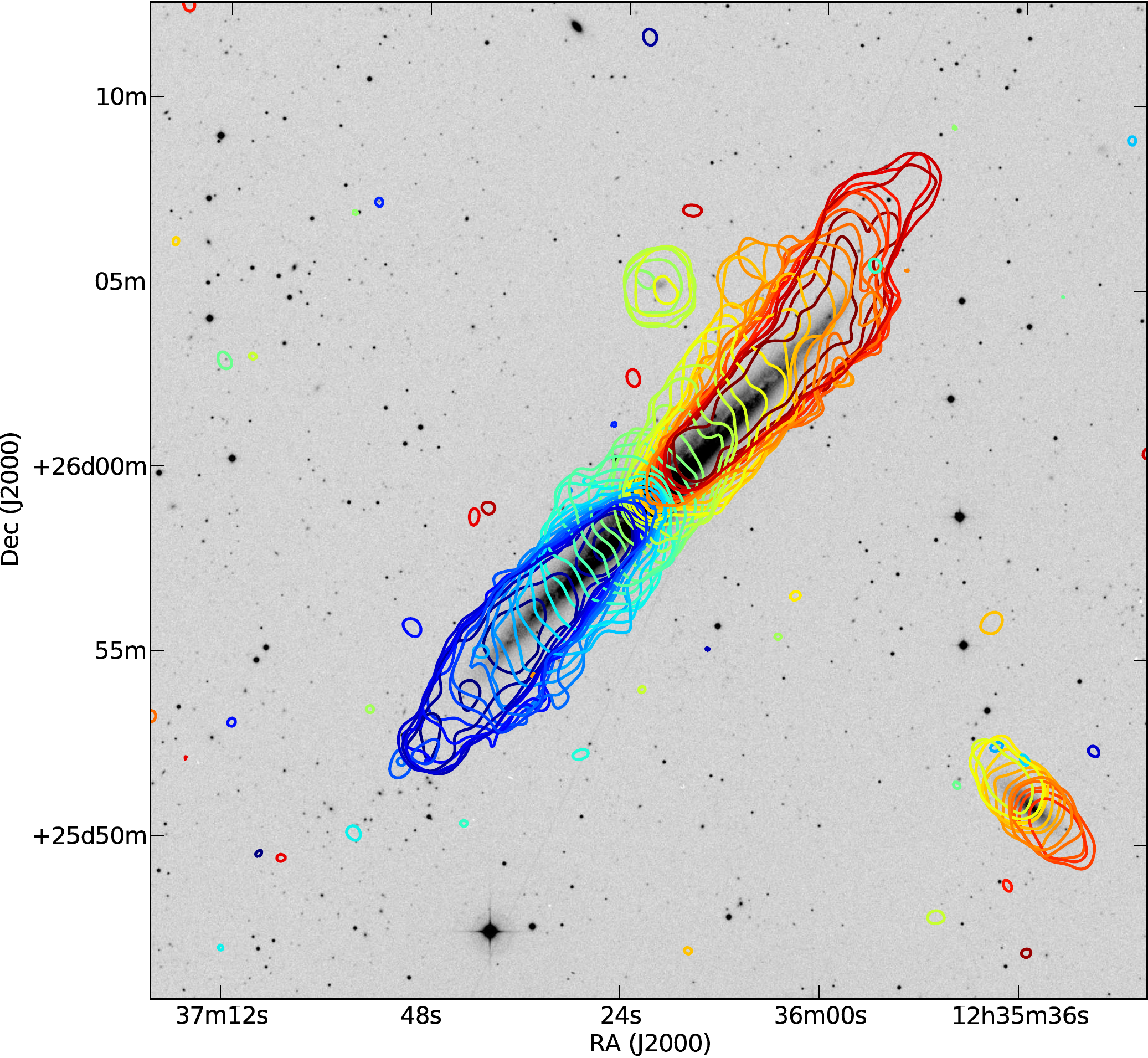}}
\caption{Overview of the HALOGAS observations of NGC~4565. The panels are as in Figure \ref{figure:u2082}. The contours in the left panel begin at $N_{HI}\,=\,2.0\times10^{19}\,\mathrm{cm^{-2}}$ and increase by powers of two. The contours in the right panel are all at $1.0\,\mathrm{mJy\,beam^{-1}}$, and are shown for a number of channels separated by $12.4\,\mathrm{km\,s^{-1}}$, starting at $945\,\mathrm{km\,s^{-1}}$ (dark blue) and ranging upwards to $1489\,\mathrm{km\,s^{-1}}$ (dark red).}
\label{figure:n4565}
\end{figure*}

\begin{figure}
\resizebox{\hsize}{!}{\includegraphics{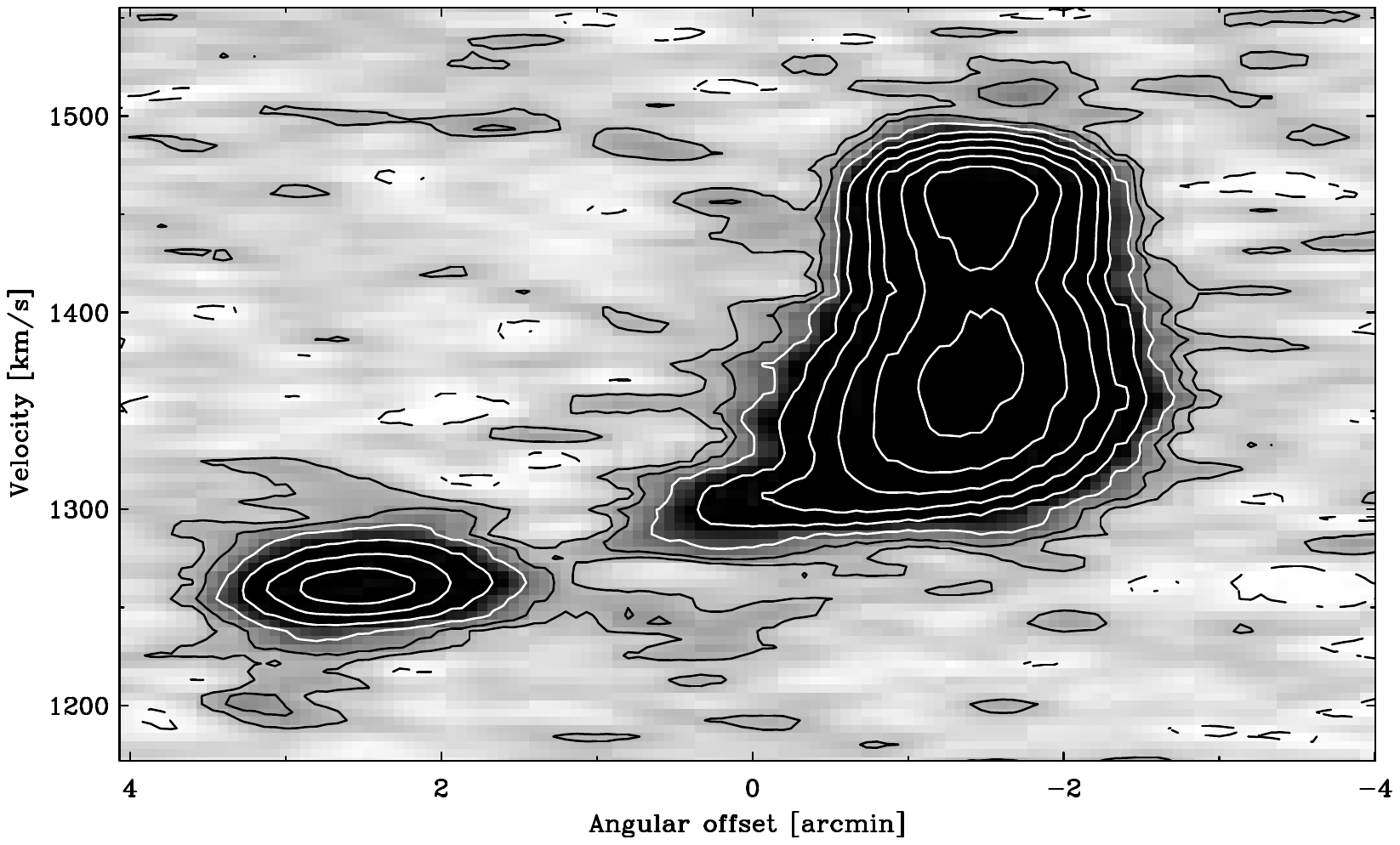}}
\caption{As in Figure \ref{figure:pvdiagram_u2082}, but for NGC~4565. The line in Figure \ref{figure:n4565} indicates the orientation of the PV slice shown here. The northeast end of the slice is to the left. The lowest contour level (solid and dashed) is $\pm350\,\mu\mathrm{Jy\,beam^{-1}}$. At the distance listed in Table \ref{table:sample}, $1^\prime$ corresponds to a linear scale of $2.82\,\mathrm{kpc}$.}
\label{figure:pvdiagram_n4565}
\end{figure}

The distribution of \HI\ emission is shown in Figure \ref{figure:n4565}. The total \HI\ mass detected in NGC~4565 itself is $7.43\times10^9\,M_{\odot}$. The galaxy is obviously warped: this can be seen in the total \HI\ map (particularly obvious on the north end of the disk) and also in the first-moment image contours. NGC~4562 has a regularly rotating \HI\ disk with a total \HI\ mass of $1.82\times10^8\,M_{\odot}$. IC~3571 is also detected in \HI\ ($M_{\mathrm{HI}}\,=\,4.18\times10^7\,M_{\odot}$), and the \HI\ data show clear signs of a minor interaction between it and NGC~4565.

NGC~4565 has a moderately high star formation rate when compared to the rest of the HALOGAS sample. The most similar HALOGAS sample galaxy (with existing observations to date) is NGC~891. There are, however, several key differences between these two galaxies. First, the SFR of NGC~4565 is only about one quarter of that of NGC~891. Both galaxies have high rotation speeds (but that of NGC~4565 is 15\% higher -- it has the highest rotation speed of all the HALOGAS sample galaxies). The distances to NGC~4565 and NGC~891, and their optical diameters, indicate that NGC~4565 is larger in linear dimensions.

\begin{figure}
\resizebox{\hsize}{!}{\includegraphics{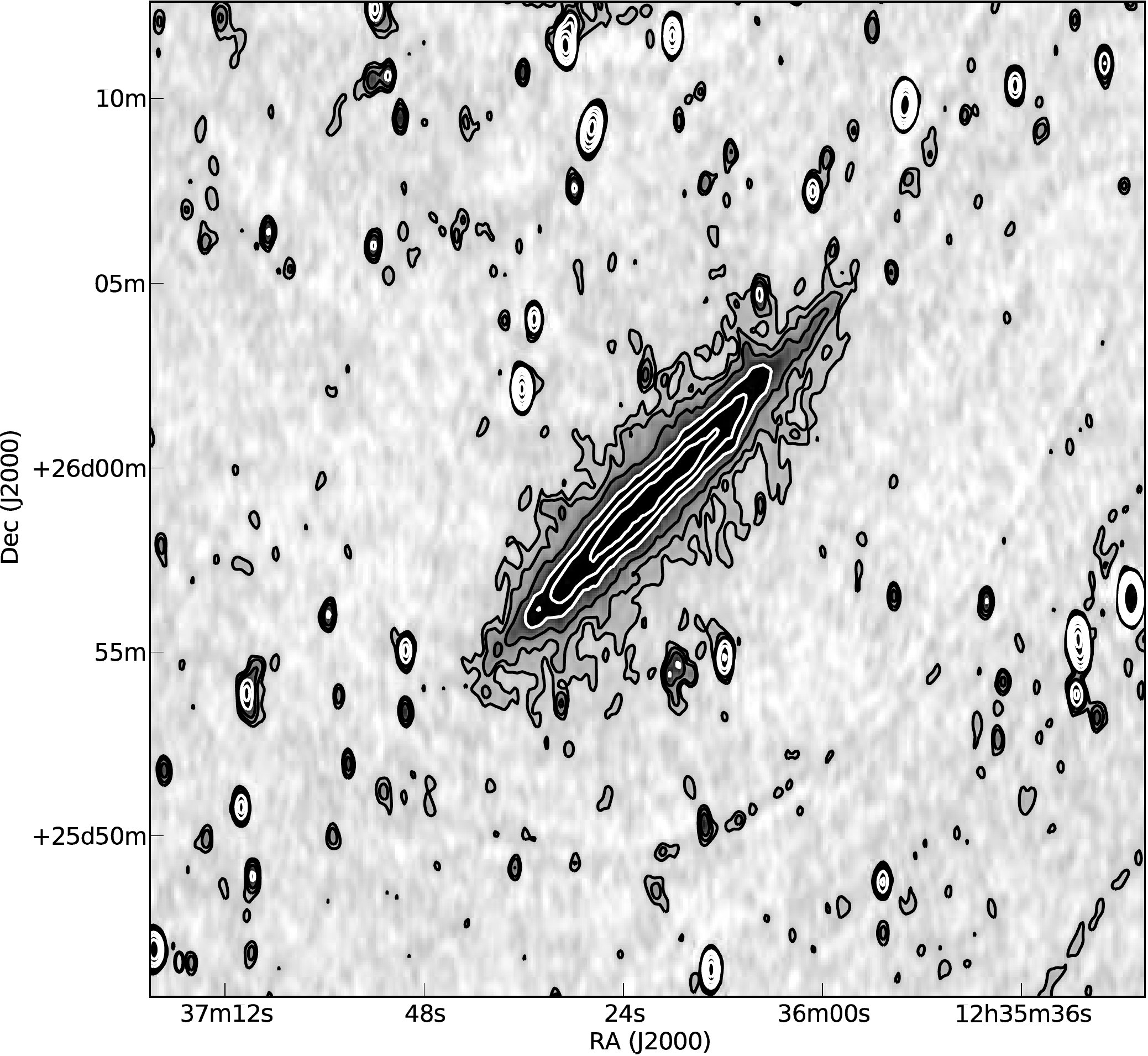}}
\caption{Radio continuum image of NGC~4565. The contour levels begin at $54\,\mu\mathrm{Jy\,beam^{-1}}\,(\approx3\sigma)$ and increase by powers of two.}
\label{figure:n4565continuum}
\end{figure}

It is apparent from Figure \ref{figure:n4565} that the \HI\ distribution in this galaxy is far less vertically extended than that of NGC~891 \citep[compare with Figure 1 of][]{oosterloo_etal_2007}, even though our new observations have comparable sensitivity to the NGC~891 work. Although the disk of NGC~4565 is clearly somewhat warped \citep[and importantly, also partially along the line of sight;][]{rupen_1991}, we can still derive a reasonable {\it upper limit} to the vertical extent, of approximately 5 kpc \citep[using $\mathrm{PA}\,=\,315^\circ$;][]{dahlem_etal_2005}. Meanwhile, the \HI\ halo of NGC~891 extends up to more than 10 kpc -- a distinct difference. Previous work has also determined that there is a lack of ionized gas in the halo of NGC~4565 \citep[see][]{rand_etal_1992}. A final difference between the two galaxies is that NGC~891 has a extremely bright and vertically extended radio continuum halo \citep[e.g.,][]{dahlem_etal_1994}. We find that NGC~4565 also has a vertically extended, and to our knowledge previously undetected at these frequencies, radio continuum halo (see Figure \ref{figure:n4565continuum}). However, it does not seem to be as extended (the continuum halo in NGC~4565 reaches up to $2-3$ kpc, compared to 10 kpc in NGC~891), nor is it as bright, as the halo in NGC~891. Studies of the continuum emission (and its polarization) in NGC~4565 are described by \citet{hummel_etal_1984} and \citet{sukumar_allen_1991}. Further discussion regarding the differences between NGC~4565 and NGC~891 is provided in \S\,\ref{section:discussion}.

\section{Discussion}\label{section:discussion}

The HALOGAS pilot survey, while still limited in the number of targets, nevertheless covers a substantial range of galactic properties. From the perspective of environmental properties, we have an apparently isolated galaxy (UGC~2082), two targets that have small, gas-rich companions (NGC~925 and NGC~4565), and one galaxy in an interacting pair (NGC~672). We also sample about two orders of magnitude in star formation rate, and nearly a factor of three in rotation speed, with this pilot survey. While it is clear that none of the pilot survey galaxies have an \HI\ halo of the scale seen in NGC~891, we find that all of the pilot survey targets show varying indications of cold gas in their halos.

The lack of a prominent halo in UGC~2082 may not be a surprising result: the star formation rate is low, and the galaxy is isolated. Apart from the possibility of significant primordial accretion, a ready source for creating a gaseous halo does not seem to be present in this target. Nonetheless, we find several distinct extraplanar features (these are noted in \S\,\ref{subsection:u2082} and are visible in Figures \ref{figure:u2082} and \ref{figure:pvdiagram_u2082}). These clouds may be a sign of cold gas accretion.

Two of the pilot galaxies that are located in denser environments have a much less regular \HI\ structure. Clearly, the case of NGC~672 is a complicated one -- the strong interaction with its companion IC~1727 has spread \HI\ emission throughout the system. This is a dramatic case of interaction-induced accretion. Another interacting system but at a later stage, NGC~925 seems to have interacted with a gas-rich companion sometime in its recent past, resulting in stripped material surrounding the main disk. Whether the culprit is the faint blob at the southern edge of the galaxy, the object found to the north (as described in \S\,\ref{subsection:n0925}), or some other entity, is not yet clear. NGC~925 also seems to possess \HI\ at lagging velocities with respect to the main disk. The anomalous component in NGC~925 has a similar mass and kinematic lag when compared to the same feature in NGC~2403.

Another target to compare with existing observations is the large edge-on NGC~4565, which does not show evidence of a prominent halo component. Can this lack be explained? In comparison with NGC~891, the differences in the global parameters of the two galaxies all tend to be in the sense in which one would indeed expect a larger halo in NGC~891, if gaseous halos are induced primarily by star formation-driven galactic fountain flows. Under that assumption, the most important properties of a galaxy for determining whether a halo is formed should be (1) the SF energy injection rate \citep[per unit disk area; see e.g.][]{dahlem_etal_1995,rossa_dettmar_2003a}, and (2) the gravitational potential (traced by $v_{\mathrm{rot}}^2$). With NGC~891's larger SFR, smaller disk area (at the distances listed in Table \ref{table:sample}), and lower rotational velocity, it would seem that it should indeed be more efficient in maintaining a gaseous halo. Per unit disk area, more star formation energy is available, and gravity is less effective at preventing material from moving away from the disk midplane. That NGC~925 and NGC~2403 have similar anomalous gas features, as well as similar SFR and mass, would also be consistent with such a picture of gaseous halo formation.

Do the other HALOGAS targets follow the same trend with the global galaxy parameters? Though it is intriguing, the number of galaxies which have been observed with enough sensitivity to distinguish a population of halo gas remains rather low, so that general statements are still uncertain. This initial result from our pilot survey underlines the need for a suitably large survey, in order to trace a range of variables in the population of nearby galaxies. The full HALOGAS survey will provide such a large sample.

\section{Multi-wavelength supplements}\label{section:multiwavelength}

For a full census of the contents of the survey galaxies, our \HI\ observations must be supplemented with high-quality, sensitive observations tracing the distribution and kinematics of stars, dust, and other gas phases. With that in mind, the WSRT HALOGAS program is being supplemented in several ways.

Since minor mergers are thought to be one of the major channels for gas accretion, stellar- and gas accretion are profoundly linked and cannot be studied independently. We therefore obtained observing time with the Isaac Newton Telescope (INT; using the Wide Field Camera) in order to complement the deep HALOGAS \HI\ observations with a photometric survey suited to detect the stellar remnants of former accretion events. The INT observations for a subset of the HALOGAS sample have been obtained, and the data will be presented in a forthcoming companion paper. This optical observing program is named HALOSTARS.

The 3.5m ARC telescope at Apache Point Observatory is being used for kinematic observations of the ionized gas in a sample which overlaps the edge-on subsample of HALOGAS, using a novel multi-slit technique that allows for simultaneous H$\alpha$ spectra of 11 parallel long slits, spaced $22\arcsec$ apart over a $4.5\arcmin\times4.5\arcmin$ field of view. Together this set of long slits can be used to form a velocity field. So far, data have been obtained for NGC~891, NGC~3044 (not a HALOGAS target), NGC~4244, NGC~4565, and NGC~4631. The observations will continue through the next several observing seasons. These optical observations will allow us to test whether both the ionized and neutral phases share the same origin and kinematics \citep[as has been previously investigated in NGC~891;][]{heald_etal_2006}. Moreover, the ionized component is more tightly coupled to the radial extent of the star forming disk than is the neutral, likely resulting in little to no contamination from warped outer disk gas complicating the analysis of extraplanar emission. This provides an important check on the \HI\ modeling results.

We have received 102~ksec on the Galaxy Evolution Explorer \citep[GALEX;][]{martin_etal_2005} satellite to observe 8 galaxies, of which 6 are drawn from the HALOGAS sample and 2 from a complementary Very Large Array (VLA) \HI\ program. In addition, 3 HALOGAS targets are already being observed as part of the GALEX Deep Galaxy Survey (DGS) survey, while NGC~2403 and NGC~891 also have existing deep archival GALEX exposures from other programs. We were granted observations in both GALEX bands, but unfortunately the FUV channel onboard GALEX is now unavailable, so that we will only obtain NUV images. The purposes of the GALEX observations are ({\it i}) to search for scattered light from dust co-located with the gaseous halo in edge-on systems, which would help elucidate the origin of this material; and ({\it ii}) to perform a highly sensitive search for star formation in the outer disks and halo regions in all of the galaxies, allowing us to study the correlation between gas accretion and outer disk star formation.

Finally, we also plan on obtaining wide-field UBRH$\alpha$ imaging in support of the GALEX and WSRT observations to search for outer disk and halo star formation, and for companions to determine to what extent accretion of gas may be associated with gas rich faint dwarfs.

\section{Data release and availability}\label{section:datarelease}

The calibrated survey data will be made publicly available in stages as the observations, data reduction, and analysis progress. The survey webpage is located at {\tt http://www.astron.nl/halogas} and this is the primary location at which reduced data cubes, moment maps, and other data products can be found. Data releases will coincide with the publication of analysis papers. Pre-release data products may be obtained by contacting the HALOGAS team.

\begin{acknowledgements}
The Westerbork Synthesis Radio Telescope is operated by ASTRON (Netherlands Institute for Radio Astronomy) with support from the Netherlands Foundation for Scientific Research (NWO). This material is based in part upon work supported by the National Science Foundation under Grant No. AST 0908106 to RJR and AST 0908126 to RAMW. RAMW also acknowledges support from Research Corporation for this project. GG is a postdoctoral researcher of the FWO-Vlaanderen (Belgium). We acknowledge the usage of the HyperLeda database ({\tt http://leda.univ-lyon1.fr}) in the course of constructing the survey sample. This research has made use of the NASA/IPAC Extragalactic Database (NED) which is operated by the Jet Propulsion Laboratory, California Institute of Technology, under contract with the National Aeronautics and Space Administration.
\end{acknowledgements}

\bibliographystyle{aa} 
\bibliography{15938}

\end{document}